\documentclass[aps, pra, 10pt, twocolumn, superscriptaddress,floatfix]{revtex4-1}

\usepackage{amsmath,amssymb,amsfonts}
\usepackage{braket}
\usepackage[breaklinks=true,colorlinks,citecolor=blue,linkcolor=blue,urlcolor=blue]{hyperref}
\usepackage{mathtools}
\usepackage{algorithm}
\usepackage{algpseudocode}
\newtheorem{thm}{Theorem}[section]
\def \var {\Delta^2 \hat{\theta}}
\DeclarePairedDelimiter{\ceil}{\lceil}{\rceil}
\newcommand{\round}[1]{\ensuremath{\lfloor#1\rceil}}

\begin{document}
\title{Achieving Heisenberg scaling with maximally entangled states: an analytic upper bound for the attainable root mean square error} 

\author{Federico Belliardo}
\email{federico.belliardo@sns.it}
\affiliation{NEST, Scuola Normale Superiore, I-56126 Pisa,~Italy}

\author{Vittorio Giovannetti}
\email{vittorio.giovannetti@sns.it}
\affiliation{NEST, Scuola Normale Superiore and Istituto Nanoscienze-CNR, I-56126 Pisa, Italy}
\begin{abstract}
In this paper we
explore the possibility of performing Heisenberg limited quantum metrology of a phase, without any prior, by employing only maximally entangled states. Starting from the estimator introduced by Higgins et al. in New J. Phys.~{\bf 11}, 073023 (2009), the main result of this paper is to produce an analytical upper bound on the associated Mean Squared Error which is monotonically decreasing as a function of the square of the number of quantum probes used in the process. The analyzed protocol is non-adaptive and requires in principle (for distinguishable probes) only separable measurements. We explore also metrology in presence of a limitation on the entanglement size and in presence of loss.
\end{abstract}

\maketitle

\section{Introduction}
\label{sect:introduction}
Quantum metrology~\cite{Giovannetti2011, Braun2018,SCIARRINO2020} is a special sector of quantum information theory with a large variety of potential applications, spanning from probing delicate biological systems~\cite{Taylor2016} to squeezing enhanced optical interferometry~\cite{Caves1981, Demkowicz2015} and gravitational wave detection~\cite{Acernese2019, Tse2019}, alongside with magnetometry~\cite{Budker2007, Koschorreck2010, Wasilewski2010, Sewell2012, Troiani2018} and atomic clocks~\cite{Ludlow2015, Louchet2010, Kessler2014}. This last two are notable applications of atom based enhanced sensors~\cite{Degen2017, Pezze2018}, which have been found rich in uses~\cite{Bongs2019}. Arguably the most intriguing result in the field is the so called Heisenberg Scaling (HS)~\cite{Giovannetti2004,Giovannetti2006} according to which the achievable accuracy in estimating an unknown phase parameter encoded into a quantum probing system, is predicted to decrease as the inverse of the total number $N$ of probes employed in the process, overcoming the Standard Quantum Limit (SQL) $N^{-1/2}$ scaling dictated by a mere statistical arguments. This is a direct consequence of the Quantum Cram\'er-Rao (QCR) bound~\cite{QCR,QCR1} which, by maximizing the Quantum Fisher Information (QFI) of the problem upon all possible input states of the probes, gauges the ultimate susceptibility of the latter with respect to small variations of the parameter we want to estimate. Unfortunately, even without considering the technical limitations associated with the preparation of the optimal QFI input states and with the implementations of high-performing quantum readouts, translating the HS susceptibility enhancement into an effective estimation accuracy is typically not as simple as one could expect from general principles. Indeed, it turns out that any estimation procedure aimed to directly recover the value of the unknown parameter from the optimal states identified through the QFI analysis, is bound to suffer from a loophole that renders the whole strategy ineffective for metrology in the absence of prior information. Such failure can be ultimately ascribed to an extra bias term appearing in the QCR bound which doesn't go to zero in the $N$ large limit, hence compromising the HS scaling. The message here is that although optimal input probe states have maximal precision in terms of the QFI we cannot use them to estimate a totally unknown parameter by only performing measurements on such states. The underlying problem is that the QFI doesn't offer the actual achievable bound for the estimation precision, but it can rather differ a lot from it, raising the question of whether HS is reachable at all.

The works dealing with this question can be roughly divided between two approaches. The first one, concerns the determination of the state that minimizes directly the actual Root Mean Square Error (RMSE) of the estimator or the associated Holevo variance~\cite{HOLEVOVARIANCE}. In particular, in the case of a two mode interferometer aimed to recover an unknown optical phase term $\theta$, Berry and Wiseman~\cite{Berry2000} computed the optimal state of $N$ photons (the so called {\it sine state} $|\psi_{\text{sin}}\rangle$), which is equivalent~\cite{Berry2009} to the state computed by Hayashi~\cite{Hayashi2010}. A covariant measurement~\cite{Holevo} on $|\psi_{\text{sin}}\rangle$ (after the encoding of $\theta$) allows the extraction of the phase with an asymptotic precision of $\pi/N$, being it the best performance achievable~\cite{Gorecki2020}. This photonic state can be transformed in a state of distinguishable (qubit-like) probes with the same statistical properties~\cite{Hayashi2010}, yet it is worth stressing that it has no multipass counterpart where one trades the number of employed probes with an equivalent number of multiple imprinting of the phase into the state of a single probe~\cite{Giovannetti2006}-- a trick that in some cases allows one to simplify the implementation of the metrological scheme~\cite{Maccone2013, Boixo2012}. Some experiments realizing the sine state for small $N$ have also been performed~\cite{Daryanoosh2018}. The optimal covariant measurement is hard to realize experimentally with entangling operations but it can be well approximated by single photon adaptive measurements~\cite{Berry2000, Hentschel2010, Peng2020}. This approximations come though with no analytical study on the achievability of HS, nevertheless they work well numerically. The second approach relays on properly splitting the total number of available resources (say the total number of probes employed in the process or the total number of parameter imprinting steps in the multipass formulation of the problem) into ordered groups of increasing complexity, in an effort to progressively reduce the uncertainty of the unknown phase. In particular, taking inspiration from the Quantum Phase Estimation Algorithm (QPEA)~\cite{Nielsen2010, Kitaev1995, Cleve1998} which in its basic form doesn't give HS~\cite{Kaftal2014}, in Ref.~\cite{Higgins2007, Berry2009, Wiseman2009-2} numerical evidence were presented in support of the fact that such result can instead be achieved by testing the collection of groups through a properly crafted sequence of adaptive measurements -- see also Ref.~\cite{Suzuki2020} where, using the resource distributions of the modified QPEA~\cite{Higgins2007}, an HS for the amplitude estimation problem was derived. A fully independent analysis of the loophole problem in the adaptive measurement scenario has also been carried out in Ref.~\cite{Boixo2008} where, approximating with Gaussian curves the probability distributions of measurement outcomes and estimators, Boixo and Somma managed to restrict step by step the confidence interval of a Bayesian phase estimator in such a way to deliver the HS. A further progress in the problem was finally made by Higgins {\it et al.} in Ref.~\cite{Higgins2009} and by Kimmel {\it et al.} in the followup works~\cite{Kimmel2015, Rudinger2017}: in these papers it was presented an analytical proof that, via a proper management of the resource splitting, one can force the Holevo variance~\cite{Higgins2009} and the RMSE~\cite{Kimmel2015, Rudinger2017} of the phase estimation problem to reach the HS even without resorting to adaptive measurements, but only relaying on a clever post-processing of the acquired data. 

A first aim of our manuscript is red to present a thoughtful review of the protocol used in Refs.~\cite{Higgins2009, Kimmel2015, Rudinger2017}, giving a detailed account of all the technicalities involved in the analysis, cleaning up some minor errors, and extending it to account for regimes where the available resources do not exactly match the splitting conditions implicitly assumed in the scheme. The final result of this effort is to derive a rigorous analytical upper bound for the RMSE of the estimation process which deviates from the lower bound dictated by the HS by a multiplicative constant. In the second part of the work we analyze the performance of the protocol in some non ideal scenarios. To begin with, we discuss what happens when the entanglement size we are allowed to employ in preparing the input state of the probes (or equivalently when the total number of consecutive phase imprinting rounds in the corresponding multipass description of the problem) is limited by technological reasons: under this condition we present an analytical characterization of the transition to the SQL regime, where the attainable RMSE scales inversely with the square root of the employed probes. Second we analyze how the presence of noise (represented by the mere loss of the encoded message on the probes) affects the optimal resource distribution, both in the ideal framework and in the limited entanglement case.

The material is organized as follows: in Sec.~\ref{sect:phaseEstimation} we formulate the HS phase estimation problem and explain the loophole affecting the metrological scenario with optimal input states that maximize the associated QFI functional. In Sec.~\ref{sect:algorithm} we present the phase estimation procedure, starting from the definition of the required measurements to be performed. In this section we explain how to extract the relevant information from each measurement, how to post-process it adaptively (Algorithm~\ref{alg:nonAdaptive}), and produce an upper bound on the attainable precision. In Sec.~\ref{sect:optimization} we optimize the bound with respect to the resource splitting diagram showing that the scheme operates indeed at the HS: such optimization is performed under some simplifications, which allow for a straightforward analytical treatment but neglect to use some of the probes. Then a little further achievable improvement is obtained by optimizing the redistribution of such extra resources. 
 
In Sec.~\ref{sect:external} we deal with modified strategies useful when external limitations are imposed, such as a maximum entanglement size or a loss noise. Conclusions are presented in Sec.~\ref{CONCSEC} while technical material is reported in the Appendix. In particular Appendix~\ref{app:SEPMES} clarifies the separability of the measurements employed in the procedure of Sec.~\ref{subsect:build}. Appendix~\ref{EQUIVALENCE} proofs the equivalence of conditions in Eq.~\eqref{eq:toCite} and Eq.~\eqref{eq:widehatEstimator}. Appendix~\ref{app:genericb} contains a generalization of the main bound in Eq.~\eqref{eq:last}. Appendix~\ref{app:proof} contains the proof of Theorem~\ref{thm:optimalDeltaM}. Appendix~\ref{app:smallSteps} is a clarification on the domain of validity of Eq.~\eqref{eq:optimalRMSEK} of the main text. Appendix~\ref{app:optimalK} is a side question that arises during the resource optimization in Sec.~\ref{subsect:interpolation} and Appendix~\ref{app:adaptive} defines the adaptive measurements to be used in Sec.~\ref{subsect:limitedEnt}.

\section{The problem}
\label{sect:phaseEstimation}
\begin{figure}[!t]
	\begin{center}
		\includegraphics[width=0.4\textwidth]{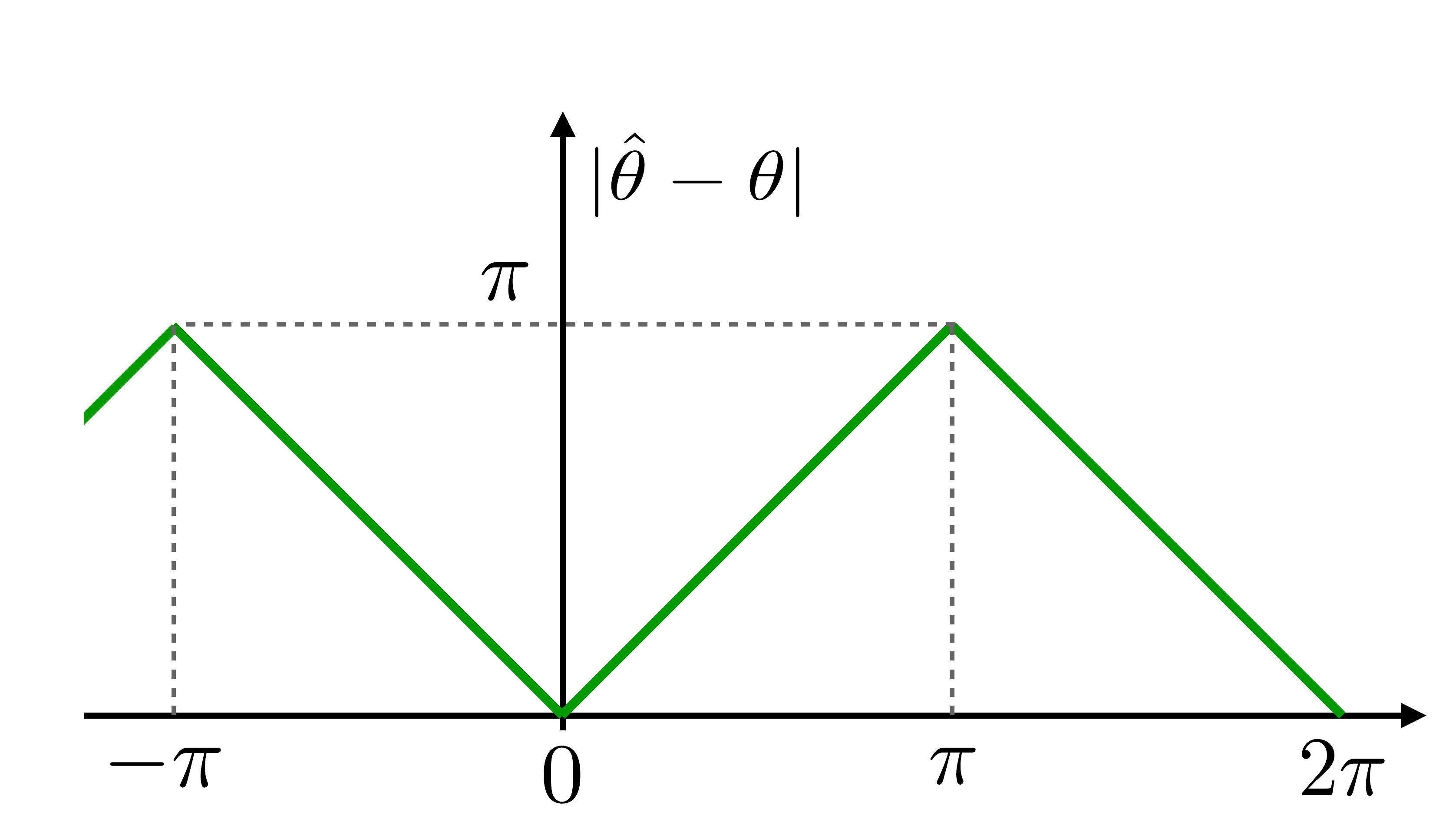}
	\end{center}
	\caption{Plot of the unit circle distance $|\hat{\theta} - \theta|$: this is piecewisely a linear ramp and has period $2\pi$. Setting $x=\hat{\theta}-\theta$ this distance can be formally expressed as $\pi - | x \bmod 2 \pi - \pi |$.}
	\label{fig:unitcircle}
\end{figure}

In our analysis we shall focus on a conventional black-box model~\cite{Giovannetti2006} where the unknown parameter $\theta$ we wish to estimate is a phase term that gets imprinted into the input state $|\psi\rangle$ of a probing quantum system via
the transformation
\begin{equation}
	|\psi\rangle \longrightarrow |\psi_\theta\rangle := U_\theta |\psi\rangle \; ,
	\label{eq:IMPRINTING} 
\end{equation} 
where
$U_\theta := e^{i \theta H}$ 
is a unitary gate generated by a fixed Hamiltonian operator $H$. In the multi-test scenario we assume to have $M$ probes initialized in a (possibly entangled) state $|\psi^{(M)}\rangle$, each evolving thanks to the application of the same black-box transformation~(\ref{eq:IMPRINTING}). The resulting output configuration
\begin{equation} 
	|\psi^{(M)}_\theta\rangle:= U_\theta^{\otimes M} |\psi^{(M)} \rangle\;,
	\label{eq:IMPRINTINGM}
\end{equation} 
is the state we can operate on to recover the value of~$\theta$. Without loss of generality we shall focus on procedures that produce an estimate $\hat{\theta}$ of $\theta$ by performing measurements on $\nu$ copies of the state $ |\psi^{(M)}_\theta\rangle$, corresponding to a total number of probes involved in the process equal to 
\begin{eqnarray} 
	N := \nu M\;.
	\label{TOTALNUMBER} 
\end{eqnarray}
Indicating with $P(\hat{\theta}|\theta)$ the conditional probability of one of such protocols, we define hence its corresponding Root Mean Square Error (RMSE) as
\begin{equation}
	\Delta \hat{\theta}:= \sqrt{E \left[|\hat{\theta} - \theta|^2\right]}\;,
	\label{RMSEdef}
\end{equation}
with $E[f(\hat{\theta})] : = \int d\hat{\theta} P(\hat{\theta}|\theta) f(\hat{\theta})$ representing the mean value of the function $f(\hat{\theta})$ of the estimator $\hat{\theta}$. The RMSE is the most important figure of merit for an estimator, as whatever other sensible definition of the estimation error (like the Holevo variance~\cite{Holevo}) is bounded by it, but the opposite is not true. Notice also that in case $\theta$ is a periodic quantity of period $2\pi$ as in the examples we shall focus in this work, the term $|\hat{\theta} - \theta|$ appearing in Eq.~(\ref{RMSEdef}) should be properly understood as the distance evaluated on the unit circle depicted in Fig.~\ref{fig:unitcircle}.
The QCR bound~\cite{QCR,QCR1} implies that, irrespectively from the selected estimation protocol, the MSE (Mean Squared Error) $\Delta \hat{\theta}^2$ is limited by the inequality
\begin{eqnarray}
	\Delta^2 \hat{\theta}= E \left[|\hat{\theta} - \theta|^2\right] \geq \frac{\left( 1 + \frac{d b_\theta}{d \theta} \right)^2}{ \nu \; {\text{QFI}(\psi^{(M)}) }} + | b_\theta |^2 \;,
	\label{eq:cramerRao}
\end{eqnarray}
first rigorously derived in~\cite{Liu2016}. In this expression $b_\theta := E[\hat{\theta}] - \theta$ is the bias of the procedure while $\text{QFI}(\psi^{(M)})$ is a functional of the imprinted state~(\ref{eq:IMPRINTINGM}), called Quantum Fisher Information (QFI)~\cite{Toth2014, Paris2009}, which gauges the sensitivity of the probe with respect to infinitesimal variations of $\theta$ and which in the present example, is given by
\begin{eqnarray} 
	\text{QFI}(\psi^{(M)}) := 4 \left( \langle (H^{(M)} )^2\rangle - \langle H^{(M)} \rangle^2 \right) \;, 
	\label{eq:QFI}
\end{eqnarray} 
where $ \langle \, \cdots \rangle$ is a short hand notation for the expectation value on $|\psi^{(M)} \rangle$, and where $H^{(M)}:= \sum_{j=1}^M H_{j}$
is the collective Hamiltonian associated with the action of $M$ black-boxes. For an estimator to be useful it must satisfy at least the asymptotic unbiasedness condition, which requires $b_\theta \rightarrow 0$ for all $\theta$ as the total number of probes used grows. Normally we also ask for $d b_\theta / d \theta \rightarrow 0$ and under such hypothesis $d b_\theta / d \theta$ gives a sub-leading term in Eq.~\eqref{eq:cramerRao}. In many cases the bias of an estimator scales as $b_\theta \propto 1/\nu$, so that also the $ b_\theta^2$ term is sub-leading when $\nu \rightarrow \infty$ (we shall see however that this term may become a problem if we try to perform Heisenberg scaling metrology with $\nu = \mathcal{O} (1)$). Assuming all these conditions Eq.~\eqref{eq:cramerRao} can hence be reduced to 
\begin{eqnarray}
	\Delta^2 \hat{\theta} \geq \frac{1}{ \nu \; {\text{QFI}(\psi^{(M)}) }} \;,
	\label{eq:QCR} 
\end{eqnarray}
which is the starting point to derive the HS~\cite{Giovannetti2004}. First of all one notices that, setting the maximum spectral gap of $H$ equal to 1 for the sake of simplicity, the maximum of \eqref{eq:QFI} is easily computed as $\text{QFI}_{\max}:= M^2$ and is obtained by taking as probe an equally weighted superposition of the minimum and maximum energy eigenvectors~\cite{Giovannetti2006} of the generator $H^{(M)}$, i.e. a GHZ-like state of the form
\begin{eqnarray}
	|{\rm GHZ}^{(M)}\rangle :=(|0\rangle^{\otimes M} + |1\rangle^{\otimes M}) /\sqrt{2} \;.
	\label{eq:GHZdef}
\end{eqnarray} 
Accordingly Eq.~(\ref{eq:QCR}) yields the following ultimate limit for $\Delta^2 \hat{\theta}$ 
\begin{eqnarray}
	\Delta^2 \hat{\theta} \geq \frac{1}{\nu M^2}\;,
	\label{eq:explicitQCRbound}
\end{eqnarray} 
which holds for all choices of the parameters $\nu$ and $M$. 
If the size of the probe $M$ is held fixed, then the QCR is called the Standard Quantum Limit (SQL), whose scaling reads $\var \ge \frac{1}{M \; N} \propto \frac{1}{N}$.
The footprint of a quantum estimation scheme is however the HS 
\begin{eqnarray}
	\var \propto \frac{1}{N^2}\;,
	\label{HSnew}
\end{eqnarray}
that follows from Eq.~\eqref{eq:explicitQCRbound} by using a single (giant) GHZ state obtained by taking $\nu = 1$, or equivalently $M=N$. As anticipated in the introductory section, attaining the scaling (\ref{HSnew}) is challenged by the fact that, after the phase imprinting stage~(\ref{eq:IMPRINTINGM}), the associated output state is given by the vector
\begin{equation}
	|{\rm GHZ}^{(N)}_{\theta} \rangle = (|0\rangle^{\otimes N} + e^{i N \theta} |1\rangle^{\otimes N}) /\sqrt{2} \;,
	\label{GHZSTATE} 
\end{equation}
which is periodic in $\theta$ with period $2 \pi/N$. This implies that in order to exploit the data obtained by measuring $|{\rm GHZ}^{(N)}_{\theta} \rangle$ we must be able to locate $\theta$ within a range of size $\propto 1/N$, so we must already know the phase $\theta$ with HS precision~\cite{Hayashi2019}. In other words the GHZ state~(\ref{GHZSTATE}) contains no information regarding in which of the intervals $\left[ \frac{2 \pi k}{N}, \frac{2 \pi \left( k+1 \right)}{N} \right)$ for $k = 0, 1, \dots, N-1$ the phase is and can only be exploited if this information is known \textit{a priori}. Such a failure ultimately can be related to the presence of the bias term in the QCR bound of Eq.~\eqref{eq:cramerRao}, which when working with estimation procedures based on a single input state $|{\rm GHZ}^{(N)}_{\theta} \rangle$ simply doesn't go to zero: neglecting this contribution as we did when writing Eq.~(\ref{eq:QCR}) may hence introduce a finite gap between the left and right and side terms of the inequality that need to be properly accounted for, possibly resulting in an overall estimation precision that can be rather different from the one predicted by Eq.~(\ref{HSnew}). The message here is that although the GHZ-like states (\ref{eq:GHZdef}) have maximal sensitivity in terms of the QFI, there is no guarantee that we can use it to estimate a totally unknown parameter $\theta$ with measurements on a single copy of one of them.

It is finally worth mentioning that the above analysis can be exactly reproduced in the multipass version of the problem where the vector~$|\psi^{(M)}_{\theta}\rangle$ of Eq.~(\ref{eq:IMPRINTINGM}) get replaced by $|\psi_{M\theta}\rangle= U_{M \theta} |\psi \rangle$ obtained by forcing the input state $|\psi\rangle$ of a single probe to $M$ consecutive imprinting stages~(\ref{eq:IMPRINTING}). Also in this case the ultimate lower bound for the associated MSE $\Delta^2 \hat{\theta}$ is given by Eq.~(\ref{eq:explicitQCRbound}) (obtained this time by taking as optimal input state the superposition $(|0\rangle+ |1\rangle) /\sqrt{2}$), and the possibility of reaching the HS limit~(\ref{HSnew}) is compromised by the fact that the vector $(|0\rangle+ e^{i N\theta} |1\rangle) /\sqrt{2}$ suffers by the same periodicity problem as (\ref{GHZSTATE}).
 
\section{Phase estimation algorithm}
\label{sect:algorithm}
As anticipated in the introduction, an analytical proof of the possibility of attaining the HS has been presented in Refs.~\cite{Higgins2009, Kimmel2015, Rudinger2017} by detailing an algorithm that we now review with minimal, yet not fully trivial, modifications that help in a effort to clarify some 
technicalities. The starting point of the analysis is to split the total number $N$ of available probes into an ordered collection of $K$ subgroups, each composed by a certain number of identical copies of GHZ-like states of probes. Specifically for $j=1,\cdots, K$, we shall assume the $j$-th group to contain $2\nu_j$ copies of the state 
\begin{equation}
	|{\rm GHZ}^{(M_j)} \rangle = (|0\rangle^{\otimes M_j} + |1\rangle^{\otimes M_j}) /\sqrt{2} \;,
	\label{GHZSTATEj} 
\end{equation}
with $\nu_j$ and $M_j$ fulfilling the constraint
\begin{eqnarray} 
	N =2\sum_{j=1}^K \nu_j M_j\;.
	\label{TOTALRES} 
\end{eqnarray} 
As a result the $N$ probes input state we assume in our model writes explicitly as 
\begin{equation}
	\ket{\psi_{\text{alg}}^{(N)}} :=\bigotimes_{j=1}^K |{\rm GHZ}^{(M_j)} \rangle^{\otimes {2\nu_j} } \;,
	\label{eq:wholeInput}
\end{equation}
and admits a QFI value equal to 
\begin{equation} 
	\text{QFI}(\psi_{\text{alg}}^{(N)}) = 2 \sum_{j=1}^{N} {\nu_j} M_j^2 \;. 
	\label{QFIALG} 
\end{equation}
After being imprinted via the process~(\ref{eq:IMPRINTINGM}), the GHZ-like states of each subgroup are measured independently in a non-adaptive fashion (see Sec.~\ref{subsect:build}) yielding $K$ random outcomes that are hence later properly post-processed (see Sec.~\ref{subsect:presentationAlgorithm}) in order to produce the estimated value $\hat{\theta}$ of the parameter $\theta$. The possibility of reaching the HS following this approach will be presented in Sec.~\ref{sect:optimization} by performing an explicit optimization with respect to the choices of the partitioning parameters entering in the resource decomposition (\ref{TOTALRES}).

\subsection{Measuring each GHZ-like state}
\label{subsect:build} 
\begin{figure}[!t]
	\begin{center}
		\includegraphics[width=0.35\textwidth]{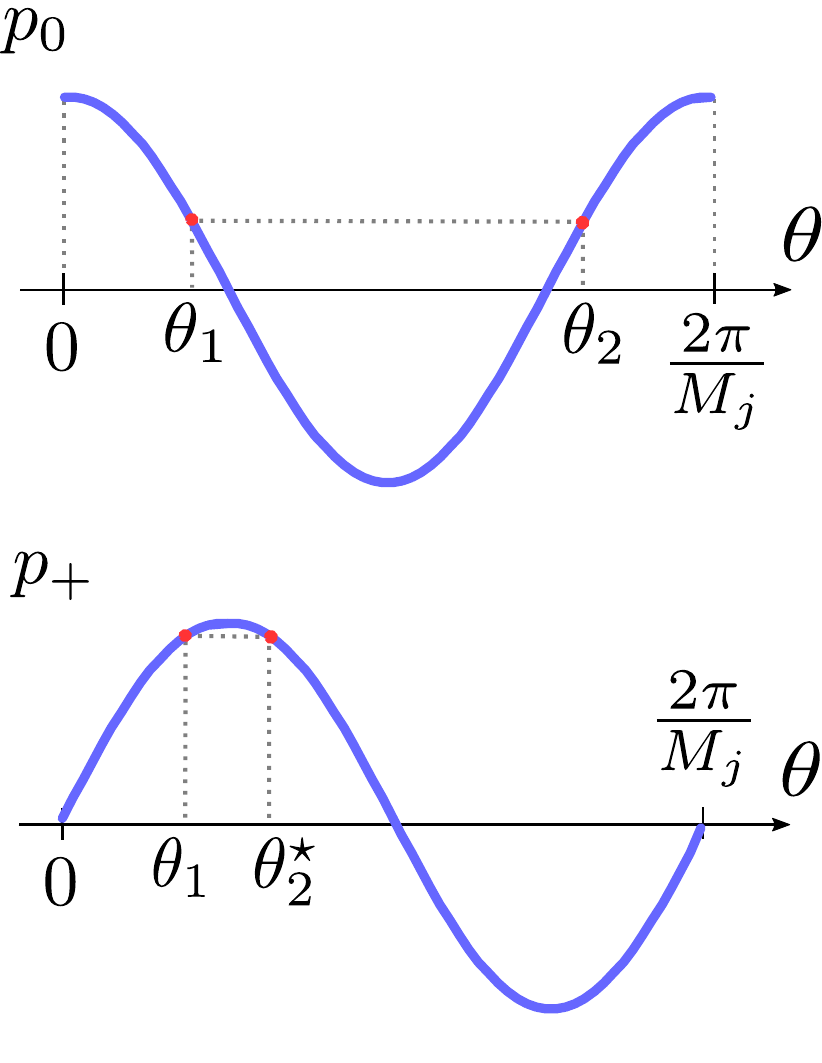}
	\end{center}
	\caption{Example of the accidental degeneracy affecting the probability~\eqref{eq:adaptiveEstimatorProbability} in the $[0,2\pi/M_j)$ interval, and its removal thanks to the interplay with probability~(\ref{eq:adaptiveEstimatorProbability+}). Two angles $\theta_1$ and $\theta_2$ correspond to the same probability value $p_0$. To lift the degeneracy we estimate also the value of $p_+$, which gives $\theta_1$ and $\theta_2^\star \neq \theta_2$ as corresponding angles. So we can identify $\theta_1$ as the angle from which $p_0$ and $p_+$ have been generated. The resolution of the ambiguity is automatic when the estimator in Eq.~\eqref{eq:estimator} is used.}
	\label{fig:doubleTheta}
\end{figure}

Here we describe the measurements we perform on each maximally entangled state of the $j$-th subgroup, which according to our construction contains $2 \nu_j$ copies of the state $|{\rm GHZ}^{(M_j)} \rangle$ of Eq.~(\ref{GHZSTATEj}). We start by noticing that given the imprinted version of such state, i.e. the vector
\begin{equation}
		|{\rm GHZ}^{(M_j)}_{\theta} \rangle = (|0\rangle^{\otimes M_j} + e^{i M_j \theta} |1\rangle^{\otimes M_j}) /\sqrt{2} \;,
		\label{OUTPUTPHI}
\end{equation}
the information on $\theta$ can be extracted by projecting it onto $(|0\rangle^{\otimes M_j} \pm |1\rangle^{\otimes M_j}) /\sqrt{2}$, a procedure which yields as outcome a Bernoulli variable with value $0$ or $1$ characterized by outcome probabilities 
\begin{equation}
	p_0 := \frac{1 + \cos M_j \theta}{2} \;, \qquad p_1= 1-p_0\;.
	\label{eq:adaptiveEstimatorProbability}
\end{equation}
Such outcome can be obtained by employing only local detection of the individual probes forming each maximally entangled state. This means that we can build an outcome variable distributed with the probabilities~\eqref{eq:adaptiveEstimatorProbability} without even performing entangled measures but only separable measurements on each individual system composing the state~\eqref{OUTPUTPHI}. If the probes are qubits this means performing only single qubit measurements. Unfortunately this result applies only to distinguishable probes and not for example to photons loaded in an optical mode for which we have to apply an entangled measure on each GHZ-like state (see Ref.~\cite{Bollinger1996} and the discussion presented in Appendix~\ref{app:SEPMES} for details). There is thought an issue still to be solved. If we perform only this kind of measurement, namely the one that gives for every GHZ-like state a Bernoulli variable with probabilities~\eqref{eq:adaptiveEstimatorProbability}, even after confining the phase to a specific period of size $\frac{2 \pi}{{M}_j}$, due to the accidental degeneracy associated with the functional $\theta$-dependence of the probabilities (\ref{eq:adaptiveEstimatorProbability}), two distinct values of $\theta$ will give the same statistics-- see Fig.~\ref{fig:doubleTheta}. To cope with this issue, one can resort in performing two types of measurements (called Type-$0$ and Type-$+$), one projecting a fraction of the copies of the state $|{\rm GHZ}^{(M_j)}_{\theta} \rangle$ on $(|0\rangle^{\otimes M_j} \pm |1\rangle^{\otimes M_j}) /\sqrt{2}$ as before, and the other projecting the remaining copies on $(|0\rangle^{\otimes M_j} \pm i |1\rangle^{\otimes M_j}) /\sqrt{2}$. Indicating the outcomes of the Bernoulli variable produced by the Type-$0$ measurement with the symbol $0,1$, we have that their associated probabilities are again expressed as in Eq.~(\ref{eq:adaptiveEstimatorProbability}); on the contrary indicating with $+,-$ the outcomes of the Bernoulli variable produced by the Type-$+$ measurement, we have that their probabilities are given by
\begin{equation}
	p_+ := \frac{1 + \sin M_j \theta }{2} \;, \qquad p_-= 1-p_+\;,
	\label{eq:adaptiveEstimatorProbability+}
\end{equation}
whose functional dependence on $\theta$ allows us to resolve the above mentioned accidental degeneracy of (\ref{eq:adaptiveEstimatorProbability}). Also the outcome of a Type-+ measurement can be realized by resorting only to individual detections on each probe constituting the state~\eqref{OUTPUTPHI}, if the probes are distinguishable.
In particular, repeating $\nu_j$ measurements of Type-$0$ and $\nu_j$ measurements of Type-$+$, each time burning one of the $2\nu_j$ resources, we define the observed probabilities of the process as
\begin{equation}
	f_0 := \frac{a_0}{\nu_j}, \qquad f_+ := \frac{a_+}{\nu_j} \;,
	\label{eq:obsfreq}
\end{equation}
where $a_0$ and $a_+$ represent, respectively the recorded values of $0$ and $+$ outcomes. The quantities $f_+$ and $f_0$ are (bounded) independent random variables which, due to the (Weak) Law of Large Numbers, for $\nu_j\rightarrow \infty$ converge in probability to their associated expectation values
\begin{equation} \label{PROBLIM} 
	f_{0/+} \xrightarrow{prob} \, p_{0/+} \; .
\end{equation}
It is worth stressing that the prospected measurement scheme is chosen \textit{a priori} and doesn't depend on the runtime result of the previous measurements neither on the actual value of $\theta$. This means that the measurement is non-adaptive. On the contrary the estimator $\hat{\theta}$ produced at each step will be dependent on the history of the previous constructed estimators, hence it will be adaptive. To reach the HS it will be important to gauge the resource distribution $\nu_j$ and reprocess correctly the data produced by the measurement, this last is the task of Algorithm~\ref{alg:nonAdaptive} we discuss in the next section. From the outcome of the fixed measurements we extract at each step the quantity $\widehat{M_j \theta}$, defined as:
\begin{equation}
	\widehat{M_j \theta} := \text{atan2} \left( 2 f_+ -1, 2 f_0 -1 \right) \in [ 0, 2 \pi ) \;,
	\label{eq:estimator}
\end{equation}
where atan2 is the 2-arguments arctangent casted in $[ 0, 2 \pi )$. Notice also that the estimator defined in Eq.~\eqref{eq:estimator} is consistent: indeed since $\widehat{M_j \theta}$ is a continuous function of $f_{0, +}$, from Eq.~(\ref{PROBLIM}) it follows that it will converges in probability to the correct value, i.e. 
\begin{equation}
	\widehat{M_j \theta} \xrightarrow{prob} \text{atan2} \left( 2 f_+ -1, 2 f_0 -1 \right) = M_j \theta \bmod 2 \pi \, .
\end{equation}
The above convergence holds in the limit $\nu_j \rightarrow \infty$, however in this reviewed algorithm the typical number of repetitions $\nu_j$ is exponentially smaller than the total amount of resources used. We will see indeed that the non-asymptotic proprieties of the estimator, characterizing the small $\nu_j$ regime, play here a fundamental role in the achievability of the HS. The purpose of Algorithm~\ref{alg:nonAdaptive} is to distill from the $\widehat{M_j \theta}$s a proper estimator $\hat{\theta}$ of the phase $\theta$. It is also important to stress that in the analysis of the performances of the algorithm we will not be much interested in the MSE of $\widehat{M_j \theta}$ but rather in bounding the probability of it missing the target by far.

\subsection{Constructing the estimator}
\label{subsect:presentationAlgorithm}
The procedure that ultimately will lead us to the estimation of $\theta$ with HS precision is summarized in Algorithm~\ref{alg:nonAdaptive}. As explicitly stated in line 3 of the procedure, we shall work under the assumption that, starting from $M_1=1$, the size of the maximally entangled states (\ref{GHZSTATEj}) double with the index $j$ of the subgroup, i.e.
\begin{eqnarray} 
	M_j = 2^{j-1} \qquad \forall j=1,\cdots, K,
	\label{doubling}
\end{eqnarray} 
with the aim of using these resources to reduce by a constant shrinking factor $1/2$ the uncertainty on $\theta$ at each new step of the process, by identifying $\theta$ in a confidence interval of size $\frac{2 \pi}{3 \cdot 2^{j-1}}$ (we refer the reader to Appendix~\ref{app:genericb} for a detailed discussion on the constraints that apply when using different choices for the $M_j$). Initially the prior on the phase $\theta$ is flat, implying a complete uncertainty on the full interval $[0,2\pi)$. By using $\nu_1$ copies of a single probe ($M_1 = 1$) we try to locate the phase $\theta$ (probabilistically) in a range which is $1/3$ of the original one (only in the first step, in all the others the shrinking factor is $1/2$), i.e. having size equal to $2 \pi/3$: accordingly, at the end of this step, with a confidence that we shall evaluate in the following, we now know that $\theta \in ( \hat{\theta} - \pi/3, \hat{\theta} + \pi/3)$. Then we employ $\nu_2$ states of size $M_2=2$ and compute the quantity $\widehat{ M_2 \theta }$ as dictated in Eq.~(\ref{eq:estimator}). We know that the ratio $\widehat{M_2 \theta}/M_2$ gives an estimation of the position of $\theta$ inside the two equivalent periods of size $\pi$ in which the unitary circle is divided (see Fig.~\ref{fig:circle}). The two possible positions for $\theta$, namely $\hat{\xi}$ and $\hat{\xi} + \pi$, are opposite on the circle. Now our aim is to reduce the uncertainty by $1/2$ with respect to the previous step, that is we want to identify $\theta$ with precision $\pi/6$. We notice that one and only one of the intervals of size $\pi/3$ centered around $\hat{\xi}$ and $\hat{\xi} + \pi$ intersects the previously assessed range $( \hat{\theta} - \pi/3, \hat{\theta} + \pi/3 )$, this means that we can unambiguously discriminate between the different equivalent periods generated by the GHZ-like states. The procedure is carried out for all stages $j=1,\cdots, K-1$ until the maximum entanglement size is reached.
\begin{algorithm}[H]
	\caption{Phase estimation}
	\begin{algorithmic}[1]
		\State $\hat{\theta} \gets 0$
		\For {j = 1 to K}
		\State $M_j \gets 2^{j-1}$
		\State $ \left[ 0, 2 \pi \right) \ni \widehat{M_j \theta} \gets$ Estimated from measurements.
		\State $\left[ 0, \frac{2 \pi}{M_j} \right) \ni \hat{\xi} \gets \frac{\widehat{M_j \theta}}{M_j}$
		\State $m \gets \Big \lfloor \frac{2^{j-2} \hat{\theta}}{\pi} - \frac{1}{3} \Big \rfloor$
		\State $\hat{\xi} \gets m \frac{\pi}{2^{j-2}} + \hat{\xi}$
		\If {$ \hat{\theta} + \frac{1}{2} \frac{\pi}{2^{i-2}} \le \hat{\xi} < \hat{\theta} + \frac{3}{2} \frac{\pi}{2^{i-2}} $}
		\State $\hat{\theta} \gets \hat{\xi} - \frac{\pi}{2^{j-2}}$
		\ElsIf {$\hat{\theta} - \frac{3}{2} \frac{\pi}{2^{i-2}} \le \hat{\xi} < \hat{\theta} - \frac{1}{2} \frac{\pi}{2^{i-2}}$}
		\State $\hat{\theta} \gets \hat{\xi} + \frac{\pi}{2^{j-2}}$
		\Else
		\State $\hat{\theta} \gets \hat{\xi}$
		\EndIf
		\State $\hat{\theta} \gets \hat{\theta} - 2 \pi \lfloor \frac{\hat{\theta}}{2 \pi} \rfloor$
		\EndFor
	\end{algorithmic}
	\label{alg:nonAdaptive}	
\end{algorithm}

\subsection{RME evaluation}
\label{subsect:analysisAlgorithm}
Here we evaluate the RME we can reach following Algorithm~\ref{alg:nonAdaptive} presenting an upper bound which, upon a proper optimization with respect to the choices of the parameters $\nu_j$ (see next section), will lead
us to the HS. \\

\begin{figure}[!t]
	\begin{center}
		\includegraphics[width=0.48\textwidth]{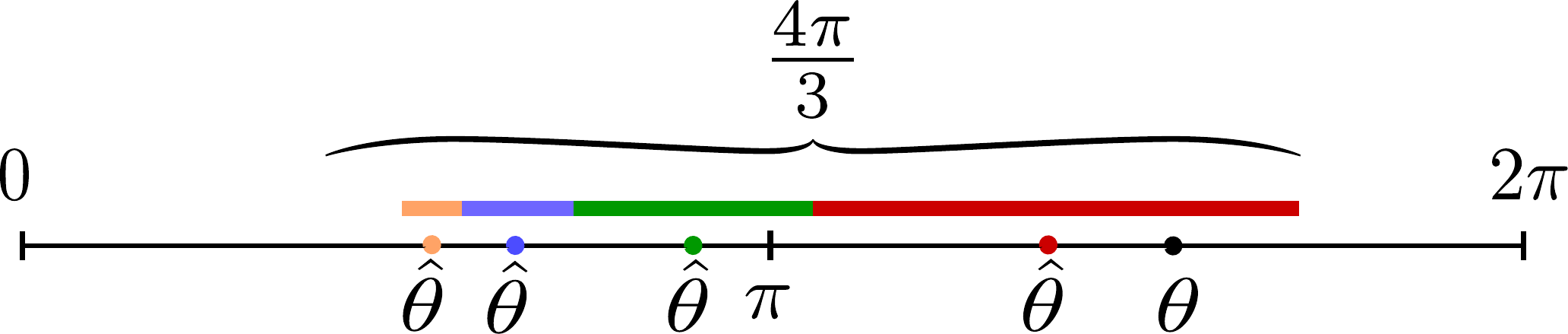}
	\end{center}
	\caption{In the worse case scenario all the estimators drift further away from $\theta$, but fortunately the total maximum error always converges.}
	\label{fig:maxDrift}
\end{figure}

Form the structure of the algorithm it is clear that to guarantee that it will return the correct result we must choose the right interval at every step. This entails that given $\hat{\theta}$ our guess for $\theta$ at the end of the $j$-th step it will fulfill the constraint
\begin{equation} 
	| \hat{\theta}- \theta | \le \frac{\pi}{ 3 \cdot 2^{j-1}} \, ,
	\label{eq:toCite}
\end{equation}
where as usual the left-hand-side is meant to indicate the distance on the unit circle (see Fig.~\ref{fig:unitcircle}) and which, as shown explicitly in Appendix~\ref{EQUIVALENCE}, can be conveniently written as
\begin{equation}
	| \widehat{M_j \theta} - M_j \theta | \le \frac{\pi}{3} \;.
	\label{eq:widehatEstimator}
\end{equation}
In view of this observation the probability of a bad estimation at the $j$-th step of the algorithm can be computed as $P \left( | \widehat{M_j \theta} - M_j \theta | \ge \frac{\pi}{3} \right)$. As it will be clear in the following, to prove that the Algorithm~\ref{alg:nonAdaptive} can reach HS it is sufficient to produce an exponential bound of the form 
\begin{equation}
	P \left( | \widehat{M_j \theta} - M_j \theta | \ge \frac{\pi}{3} \right) \le A C^{-{\nu_j}} \;,
	\label{eq:expBound}
\end{equation}
for some constants $A\geq 0$ and $C>1$. 
\begin{figure}[!t]
	\begin{center}
		\includegraphics[width=0.48\textwidth]{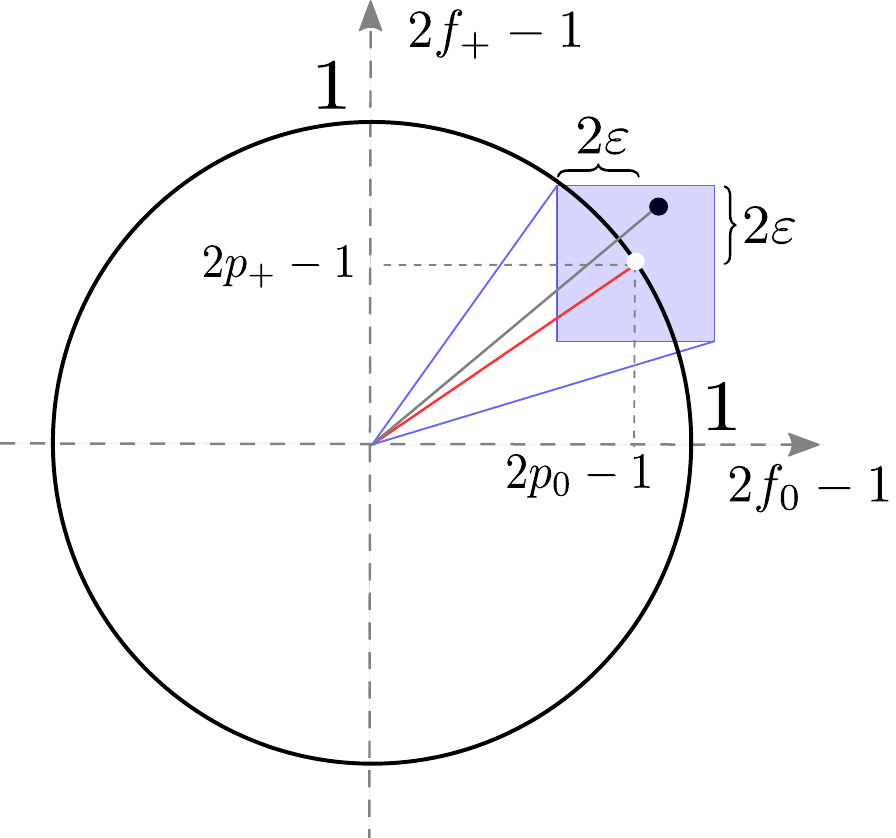}
	\end{center}
	\caption{Geometrical proof that there exists an $\varepsilon$ small enough such that when the observed frequencies $f_0$ and $f_+$ define a point $\left( 2 f_0 -1, 2 f_+ - 1 \right)$ that sits in a box of side $4 \varepsilon$ centred around $\theta$ then $| \widehat{M_j \theta} - M_j \theta | \le \frac{\pi}{3}$. The white dot in center of the blue shaded square identifies the angle $M_j \theta$ while the black dot is the measured point $\left( 2 f_0 - 1, 2 f_+ - 1 \right)$.}
	\label{fig:heisenberg}
\end{figure}
For this purpose, let us first select an $\varepsilon$ small enough such that $| f_0 - p_0 | \le \varepsilon$ and $| f_+ - p_+ | \le \varepsilon$ imply $| \widehat{M_j \theta} - M_j \theta | \le \frac{\pi}{3}$ (a choice that this is always possible as one can verify e.g. by looking at Fig.~\ref{fig:heisenberg}). Then apply the Hoeffding's bound~\cite{Hoeffding1963} on the rescaled binomial variables $f_0$ and $f_+$, obtaining
\begin{align}
	& \text{P} \left( | f_0 - p_0 | \ge \varepsilon \right) \le 2 \exp \left( - 2 \nu_j \varepsilon^2 \right) \;, \label{eq:prob1} \\ & \text{P} \left( | f_+ - p_+ | \ge \varepsilon \right) \le 2 \exp \left( - 2 \nu_j \varepsilon^2 \right) \;. \label{eq:prob2}
\end{align}
Together these inequalities imply
\begin{equation}
	\text{P} \left( | \widehat{M_j \theta} - M_j \theta | \ge \frac{\pi}{3} \right) \le 4 \exp \left( - 2 \nu_j \varepsilon^2 \right) \;,
	\label{eq:anayticalBound}
\end{equation}
and the required exponential bound has been found with $A = 4$ and $C = e^{2 \varepsilon^2} > 1$. The largest value of $\varepsilon$ that satisfies the requirements~\cite{Berg2019} is in this case $\varepsilon = \sqrt{6}/8$, which gives $C = 1.206$. We carried out a numerical evaluation of optimal $A$ and $C$ by computing exact error probabilities for each $\nu \le 80$. One hundred angles of the form $\frac{2 \pi i}{100}$ for $i = 0, 1, \dots, 99$ have been tried for every $\nu$, and the highest probability error among them has been selected. All these maximum errors are bounded as
\begin{equation}
	P \left( | \widehat{M_j \theta} - M_j \theta | \ge \frac{\pi}{3} \right) \le 0.5949 \times 1.6640^{-\nu_j} \;,
	\label{eq:expBoundNumeric}
\end{equation}
so $A = 0.5949$ and $C = 1.6640$. We stress that it is not necessary to use any numerical constant $A$ and $C$ in order to prove the Heisenberg scaling, the ones computed analytically are sufficient. Nevertheless the numerics are useful to tighten the prefactor. We are ready now to compute the MSE of the presented metrological protocol for arbitrary choices of the parameter $\nu_j$ and $K$. If no errors were made in the whole procedure, the last step, performed with states of size $2^{K-1}$, is done to reduce the range size to $\frac{2 \pi}{3 \cdot 2^{K-1}}$, so we have $| \hat{\theta} - \theta | \le \frac{\pi}{3 \cdot 2^{K-1}}$. The probability of this to happen is the product of the probabilities of all the events $| \widehat{M_j \theta} - M_j \theta | \le \frac{\pi}{3}$ for $j = 1, 2, \dots, K$. They are all independent, as each estimator $\widehat{M_j \theta}$ is a function only of the measurements outcome on the $j$-th probe bunch. Surprisingly it will be sufficient to bound the probabilities $\text{P} \left( | \widehat{M_j \theta} - M_j \theta | \le \frac{\pi}{3} \right)$ by $1$ to get HS scaling, so that the probability of getting every choice right is trivially bounded as
\begin{equation}
	\prod_{\alpha = 1}^{K} \text{P} \left( | \widehat{M_\alpha \theta} - M_\alpha \theta | \le \frac{\pi}{3} \right) \le 1 \;.
\end{equation}
Each time a new step $j$ is carried out the possible range for $\theta$ is reduced and if a wrong estimation is made all the subsequent are also wrong. We can classify all the possible estimation histories by the first wrong choice and they form disjoint classes. If the $j$-th is the first wrong choice then by definition the $(j-1)$-th choice is correct. At step $(j - 1)$-th the phase has been identified to be in a range of size $\frac{2\pi}{3 \cdot 2^{j-2}}$, but because of all the successive non reliable steps of the algorithm the phase estimator can further drift away from $\theta$. The maximum it can drift is $\frac{4\pi}{3 \cdot 2^{j-2}}$, which is obtained by summing $\frac{2 \pi}{3 \cdot 2^{j-2}} \left( 1 + \frac{1}{2} + \frac{1}{4} + \cdots \right)$, see Fig.~\ref{fig:maxDrift}. This is not a tight upper bound as the sum should contain only as many terms as steps of the algorithm yet to perform.
The probability that the first error occurs at $k = j$ is the product of the probabilities to get it right until $k = j - 1$ times the probability of doing the wrong choice at $j$, so it reads
\begin{equation}
	\text{P} \left( | \widehat{M_j \theta} - M_j \theta | \ge \frac{\pi}{3} \right) \prod_{\alpha = 1}^{j-1} \text{P} \left( | \widehat{M_\alpha \theta} - M_\alpha \theta | \le \frac{\pi}{3} \right) \;,
\end{equation}
this is bounded by $A C^{-\nu_{j}}$ as in Eq.~\eqref{eq:expBound}. Now we put everything together to find the following MSE upper bound
\begin{eqnarray}\nonumber 
	\var &=& \int (\hat{\theta} - \theta )^2 P (\hat{\theta} | \theta) \, d \hat{\theta} \\ &\le& \left( \frac{\pi}{3 \cdot 2^{K-1}} \right)^2 + \sum_{j=1}^{K} \left( \frac{8 \pi}{3 \cdot 2^{j-1} }\right)^2 A C^{-\nu_j} \nonumber \\
 	&=& \left( \frac{2 \pi}{3} \right)^2 \left( \frac{1}{4^K} + 16 \sum_{j = 1}^{K} \frac{A}{4^{j-1}} C^{-\nu_j} \right) \;.
	\label{eq:RMSEBound}
\end{eqnarray}
The first term is an upper bound on the probability of getting all the choices right times the precision squared we would have at the end. Similarly all the other terms are the product of the upper bound probability of getting the first error at $j$ times the squared upper bound on the error of the estimator at the end. The maximum error of the $j=1$ term is not precise, but its contribution to the sum will turn out to be negligible.

\section{Optimization of the resources}
\label{sect:optimization}
For all choices of the integer $K$ and of the numbers of copies $\nu_j$ which fulfill the resource constraint (\ref{TOTALRES}), the inequality~\eqref{eq:RMSEBound} provides an upper bound for the MSE attainable with the Algorithm~\ref{alg:nonAdaptive}. Aim of the present section is to show that this allows us to prove the achievability of the HS. We start in Sec.~\ref{subsect:proofHeisenberg} by employing the Lagrange multiplier technique to perform an explicit minimization of the right-hand-side of~\eqref{eq:RMSEBound} for fixed values of $N$ which ultimately leads to the inequality~(\ref{eq:last}) below. As will shall see in order to get such a clean analytical expression the approach we follow imposes a functional dependence between $N$ and $K$ that paves the way for some extra (minor) improvements which are discussed in Secs.~\ref{secB} and \ref{subsect:interpolation}.
In particular in Sec.~\ref{secB} we study the most efficient way to upgrade the maximum entanglement size employed in the process as $N$ increases, and in Sec.~\ref{subsect:interpolation} we analyze how to redistribute the extra resources that are left-over by the rigid connection between $N$ and $K$ imposed by the derivation of Eq.~(\ref{eq:last})
\subsection{Proof of Heisenberg scaling}
\label{subsect:proofHeisenberg}
Here we minimize the right-hand-side of~\eqref{eq:RMSEBound} while keeping the total number of probes constant via Lagrange multipliers. In doing so we find it useful to initially replace the integer $\nu_j$ with real variables $x_j$, and then to express the optimal solution by rounding our results to the closest integers (if needed). Under this assumption the Lagrangian of the problem reads
\begin{multline}
	\mathcal{L} := \left( \frac{2 \pi}{3} \right)^2 \left( \frac{1}{4^K} + 16 \sum_{j = 1}^{K} \frac{A}{4^{j-1}} C^{-x_j} \right) \\ - \lambda \left( 2 \sum_{j=1}^{K} x_j 2^{j-1} - N \right) \;,
	\label{eq:mainLagrangian}
\end{multline}
where we have explicitly used the fact that in our analysis $M_j = 2^{j-1}$. Imposing the stability condition with respect to variation of $x_j$, i.e. $\partial_{x_j} \mathcal{L}=0$, we hence get the identity
\begin{eqnarray}
	\lambda &=& - \left( \frac{2 \pi}{3} \right)^2 \frac{16 A \log C}{2^{3j - 2}} 2^{- x_j \log_2 C } \;,
	\label{deflambda} 
\end{eqnarray}
which, exploiting the fact that $\lambda$ cannot depend upon $j$, forces the optimal distribution of the number of copies to be close to a linear ramp (so as the states become bigger and bigger we employ less and less statistics), i.e. 
\begin{eqnarray}	
	x_j &=& \frac{3}{\log_2 C} \left( K - j \right) + x_K =\gamma \left( K - j \right) + x_K \;, \\ \nonumber && \qquad \qquad \qquad \qquad \qquad \qquad \qquad \forall j=1,\cdots K\;,
	\label{eq:linearRamp}
\end{eqnarray}
with $\gamma := \frac{3}{\log_2 C}$. Notice that the parameters $x_K$ and $K$ entering Eq.~(\ref{eq:linearRamp}) can be freely chosen under the constraint~(\ref{TOTALRES}), which formally writes
\begin{eqnarray} 
 	N &=& 2 \sum_{j = 1}^{K} \round{x_j} 2^{j-1} \;,
 	\label{nconstraint}
\end{eqnarray} 
with the rounding operation $\round{\cdot}$ introduced to compensate for the fact that Eq.~(\ref{eq:linearRamp}) will typically yields values of $\nu_j$ which are not integers. A simple analytical connection between $N$ and $K$ can be now be forced by considering the following trivial upper and lower bounds on $\round{x_j}$,
\begin{equation}
	x_j - \frac{1}{2} \le\nu_j= \round{x_j} < x_j + \frac{1}{2} \; .
	\label{eq:chainInqqnu}
\end{equation}
Replaced into (\ref{nconstraint}) this leads us to
\begin{eqnarray}
	{N^{<}_K} \leq N \leq {N^{>}_K}\;, \label{resourcebound}
\end{eqnarray} 
with
\begin{eqnarray}
	{N^{>}_K}&:=& \left( \gamma + x_K + \frac{1}{2} \right) 2^{K+1} \;, \label{eq:totalResource} \\ {N^{<}_K} &:=& \left(\gamma + x_K - \frac{1}{2} \right) 2^{K+1} \,, \label{eq:totalResourcemin}
\end{eqnarray}
which have been derived by performing the summation over $j$ and dropping negligible $\mathcal{O} \left(K \right)$ contributions in order to simplify the functional dependence upon $K$. Furthermore, replacing into~(\ref{eq:RMSEBound}) the lower bound on ${\nu_j}$ of Eq.~(\ref{eq:chainInqqnu}) 
allows us to write 
\begin{eqnarray}
	\var &\le& \left( \frac{2 \pi}{3} \right)^2 \left( \frac{1}{4^K} + \frac{64 A }{2^{3 K} C^{x_K - \frac{1}{2}}} \sum_{j=1}^{K} 2^j \right) \label{eq:total} \\ &=&\left( \frac{2 \pi}{3} \right)^2 \left( 1 + \frac{128 A}{C^{x_K-\frac{1}{2}}} \right) \frac{1}{4^K} \;.
	\label{eq:MSEbound}
\end{eqnarray}
This expression shows that the advanced steps of the estimation exponentially dominate the error. Apart from the numerical factor it closely resemble Eq.~(13) of Ref.~\cite{Higgins2009} (specifically the differences are that in Eq.~\eqref{eq:MSEbound} the size of the last error is half that of Ref.~\cite{Higgins2009}, that the size of the other contributions are increased to account for the drift of the estimator, and the presence of $x_j - \frac{1}{2}$ instead of $x_j$). To link~(\ref{eq:MSEbound}) to the total number of employed probes $N$, we can use the second inequality of Eq.~(\ref{resourcebound}) to write
\begin{equation}
	\var N^2 \le 4 \left( \frac{2 \pi}{3} \right)^2 \left( \gamma + x_K + \frac{1}{2} \right)^2 \left( 1 + \frac{128 A}{C^{x_K - \frac{1}{2}}} \right) \;,
	\label{eq:last}
\end{equation}
that explicitly proves the possibility of attaining HS precision~(\ref{HSnew}) by noticing that 
we can make $N \rightarrow \infty$ by increasing $K$ while maintaining $x_K$ constant, so that the right hand side of bound~\eqref{eq:last} remains constant.

From the numerical estimates of Eq.~\eqref{eq:expBoundNumeric} we can evaluate $\gamma = 4.0835$ and $A = 0.5949$. The prefactor of Eq.~\eqref{eq:last} can then be optimized as a function of $x_K$, revealing that
it achieves its minimum value 
\begin{equation}
	\var N^2 \le \left( 24.26 \pi \right)^2 \,,
	\label{dafa}
\end{equation}
for $x_K = 11$. The right-hand-side of (\ref{dafa}) has to be compared with $\pi^2$ which according to the recent work \cite{Gorecki2020} represents the best estimation for the multiplicative factor entering in the HS scaling~(\ref{HSnew}). It is also worth observing that this precision differs by only a factor $24.26/3.17\simeq 7.65$ from the QCR lower bound~(\ref{eq:QCR}) associated to the QFI value~(\ref{QFIALG}) of the input state~(\ref{eq:wholeInput}) of the model. Indeed in this case we have
\begin{eqnarray}\nonumber 
	 \text{QFI}^{-1}(\psi_{\text{alg}}^{(N)}) N^2 &=& \frac{N^2}{2 \sum_{j=1}^{N} {\nu_j} \left( 2^{j-1} \right)^2} \\ &\geq& \frac{({N^{<}_K})^2}{ \frac{1}{3} \left( \frac{2 \gamma}{3} + 2 x_K + 1 \right) 4^K } \label{eq:fisherPartial} \\ &=& 36 \frac{\left( \gamma + x_K - \frac{1}{2} \right)^2}{2 \gamma + 6 x_K + 3} = \left( 3.17 \pi\right)^2 \nonumber\;,
\end{eqnarray}
where the inequality follows from by inserting the upper bound of Eq.~(\ref{eq:chainInqqnu}) into the denominator and the lower bound of (\ref{resourcebound}) in the numerator, while the final expression was obtained by setting the same numerical factors we used in (\ref{dafa}). 

\subsection{Optimal upgrade of the entanglement size}
\label{secB} 
\begin{figure}[!t]
	\begin{center}
		\includegraphics[width=0.50\textwidth]{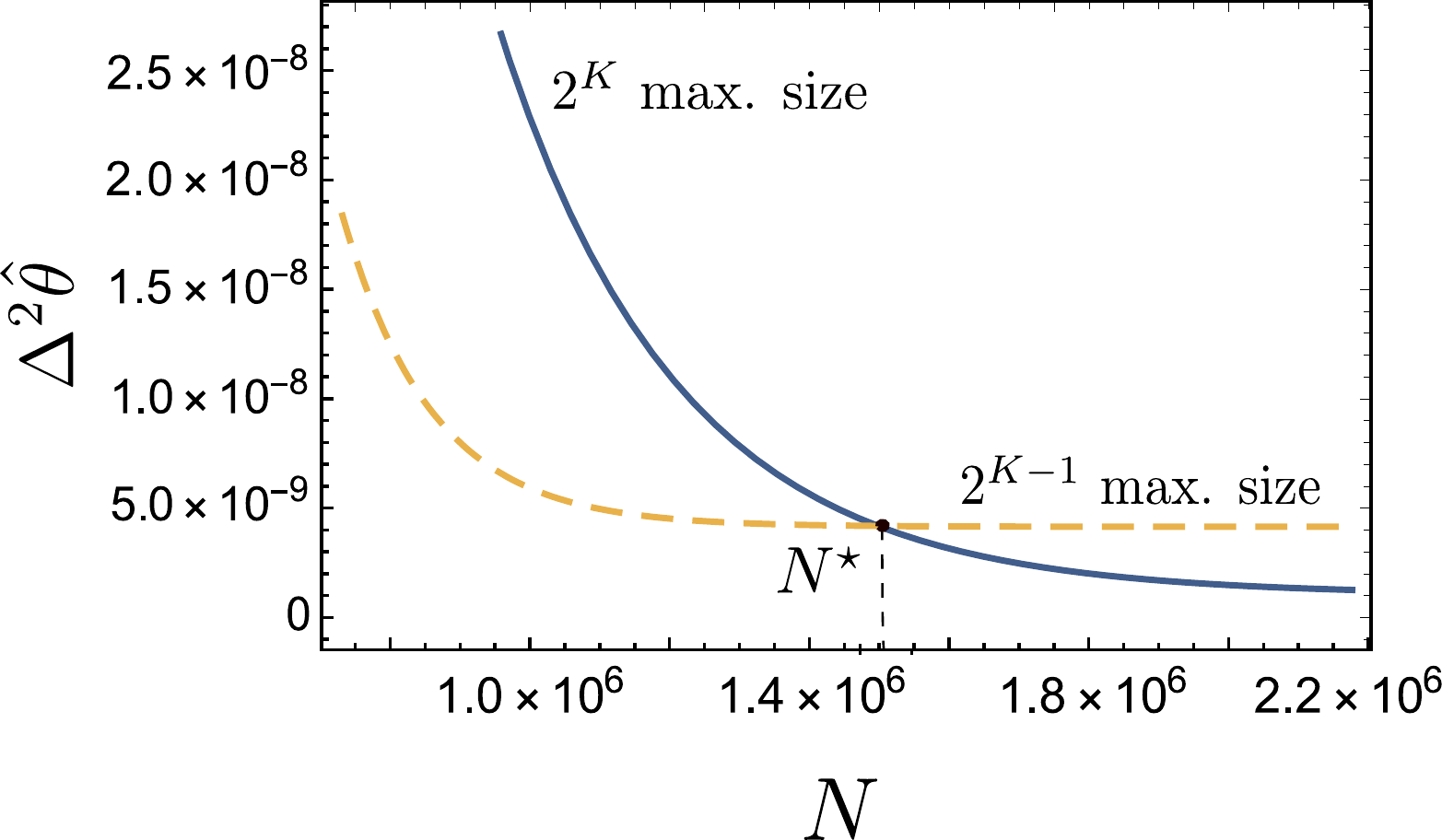}
	\end{center}
	\caption{Plot of the curves in Eq.~\eqref{eq:continousMSEbound} (dashed yellow line) and Eq.~\eqref{eq:continousMSEboundK1} (solid blue line) for $K = 15$. The dashed one corresponds to a maximally entangled size $2^{14}$ and the solid one to $2^{15}$. The values $A$ and $C$ are those of Eq.~\eqref{eq:expBoundNumeric}. The curves intersect at point $N^\star = 22.9 \cdot 2^{K+1}$.}
	\label{fig:treffPoint}
\end{figure}

\begin{figure}[!t]
	\begin{center}
		\includegraphics[width=0.48\textwidth]{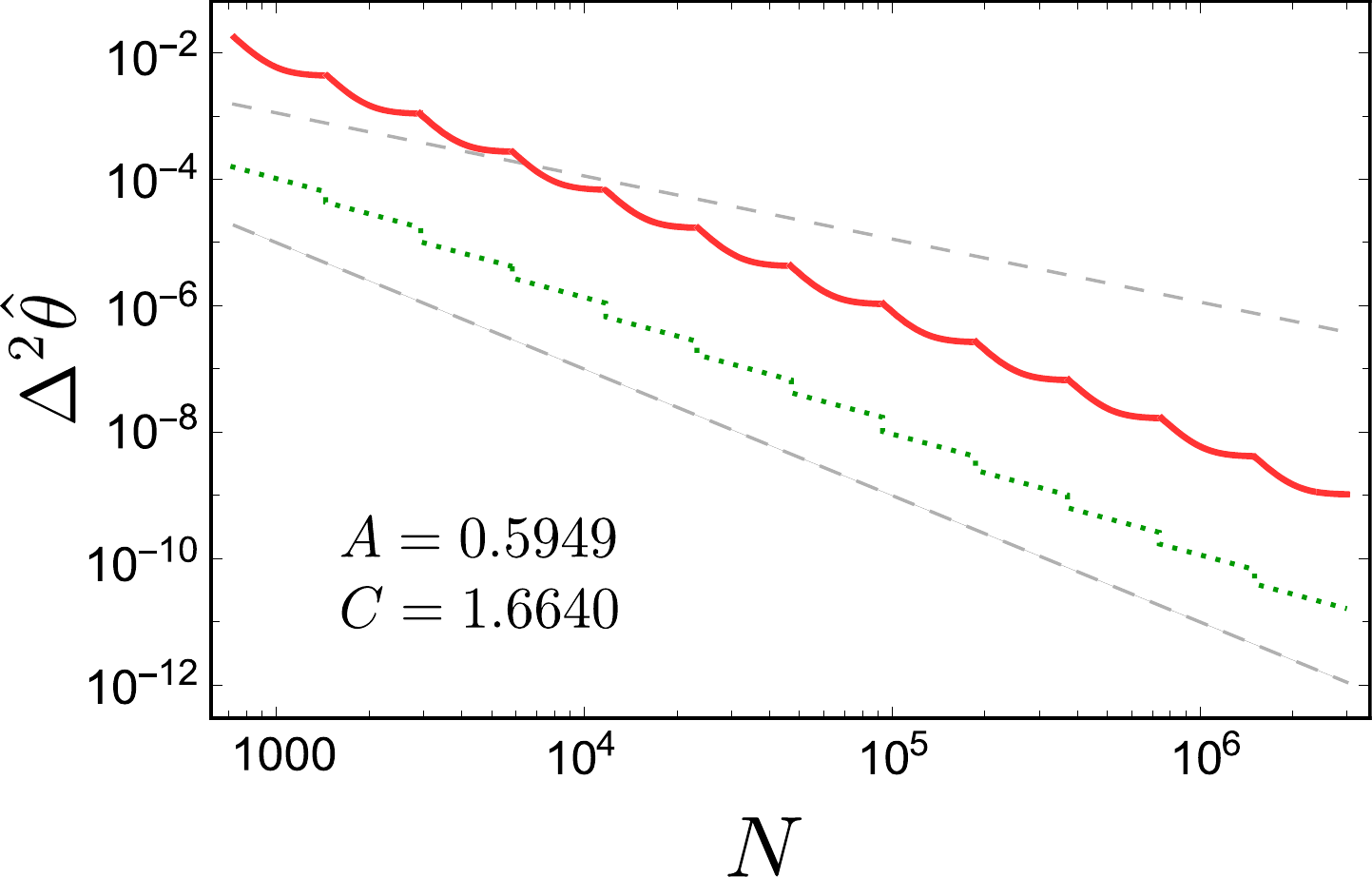}
	\end{center}
	\caption{Comparison (on a double logarithmic plot) between the Standard Quantum Limit $1/N$ (upper dashed gray line), the HS $\pi^2/N^2$ of Ref.~\cite{Gorecki2020} (lower dashed gray line), the upper bound on the MSE for the reviewed algorithm (solid red curve), obtained as a piecewise junction of Eq.~\eqref{eq:continousMSEbound} and Eq.~\eqref{eq:continousMSEboundK1} as shown in Fig.~\ref{fig:treffPoint}, and the lower bounds on $\text{QFI}^{-1}$ (dotted green curve), similarly obtained by joining Eq.~\eqref{eq:contQFI} and (\ref{eq:contQFI1}). Observe that the algorithm precision is monotonically decreasing in $N$. The numerical values of $A$ and $C$ are those of Eq.~\eqref{eq:expBoundNumeric}. Their validity conditions ($\nu_j \le 80$) are met in this plot.}
	\label{fig:finalComparison}
\end{figure}

A refinement of the inequality~(\ref{eq:last}) can be obtained by inverting Eq.~\eqref{eq:totalResource} to deduce the suitable $x_K$ corresponding to a certain ${N^{>}_K}$. Substituting such value in Eq.~\eqref{eq:MSEbound} we have
\begin{eqnarray}
	\var &\le& \left( \frac{2 \pi}{3} \right)^2 \left( 1 + \frac{128 A}{C^{\frac{{N^{>}_K}}{2^{K+1}} - \gamma -1}} \right) \frac{1}{4^{K}} \label{eq:continousMSEbound0} \\ &\le& \left( \frac{2 \pi}{3} \right)^2 \left( 1 + \frac{128 A}{C^{\frac{N}{2^{K+1}} - \gamma -1}} \right) \frac{1}{4^{K}} \, , \label{eq:continousMSEbound}
\end{eqnarray}
where the last passage was obtained by exploiting the inequality~(\ref{resourcebound}) and the monotonicity of the functional dependence of the involved term upon ${N^{>}_K}$. In Fig.~\ref{fig:treffPoint} we compare bound~\eqref{eq:continousMSEbound} and the same bound obtained after the substitution $K \rightarrow K+1$, i.e.
\begin{equation}
	\var \le \left( \frac{2 \pi}{3} \right)^2 \left( 1 + \frac{128 A}{C^{\frac{N}{2^{K+2}} - \gamma -1}} \right) \frac{1}{4^{K+1}} \, .
	\label{eq:continousMSEboundK1}
\end{equation}
The intersection between these two curves gives an idea of the location of the point $N^\star$ from which it starts to be useful to upgrade the maximum entanglement size of the input state~(\ref{eq:wholeInput}) from $2^{K-1}$ to $2^{K}$. This comparison is carried out with the numerical values given in Eq.~\eqref{eq:expBoundNumeric}. Fig.~\ref{fig:treffPoint} refers in particular to the case $K=15$ but the form of the curves is independent on $K$, this means that the position of the intersection, being $N^\star = 22.9 \cdot 2^{K+1}$ is valid $\forall K$. The value of $x_K$ corresponding to $N^\star$ is $x_K = 18.3$, while $x_{K+1}$, given by substituting $K \rightarrow K+1$ and $N^{\star}$ in Eq.~\eqref{eq:totalResource}, is $x_{K+1} = 6.87$. We conclude that while increasing the resources the optimal upgrade position is expected to be close to $\nu_K = 18$. Then we start from $\nu_K = 7$ with the upgraded maximal state size.
The upper bound on the MSE is obtained by piecewise joining the expressions in Eq.~\eqref{eq:continousMSEbound} and Eq.~\eqref{eq:continousMSEboundK1} at $N^\star$ for every $K$.
Repeating the same analysis for the QCR lower bound~(\ref{eq:QCR}) associated to the QFI value~(\ref{QFIALG}) of the input state~(\ref{eq:wholeInput}) allows us to replace Eq.~\eqref{eq:fisherPartial} with the inequality
\begin{eqnarray}
	\text{QFI}^{-1}(\psi_{\text{alg}}^{(N)}) N^2 &\geq& \frac{3N^2}{ \left( \frac{N}{2^K} - \frac{4 \gamma}{3} \right) 4^K }\;, 
	\label{eq:contQFI} 
\end{eqnarray}
where we have again inverted Eq.~\eqref{eq:totalResource} and used the upper bound~(\ref{resourcebound}), and
\begin{eqnarray} 
	\text{QFI}^{-1}(\psi_{\text{alg}}^{(N)}) N^2 
	\geq \frac{3N^2}{ \left( \frac{N}{2^{K+1}} - \frac{4 \gamma}{3} \right) 4^{K+1} }\;,
	\label{eq:contQFI1} 
\end{eqnarray}
obtained from the first one by replacing $K \rightarrow K +1$ and valid for $N \ge N^{\star}$. The resulting values are plot in Fig.~\ref{fig:finalComparison} together with the upper bound on the MSE, the reachable Heisenberg scaling $\frac{\pi^2}{N^2}$, and the SQL. According to analyzed upper bound the reachable precision of Algorithm~\ref{alg:nonAdaptive} necessarily starts to beat the SQL from $N \simeq 6 \cdot 10^3$.

\subsection{Redistribution of the extra resources}
\label{subsect:interpolation}

Given a true amount of resources $T$ we could take it as the upper bound ${N^{>}_K} = T$, then there exists a strategy with $N \le T$ that reaches an accuracy $\var$ that fulfills Eq.~\eqref{eq:continousMSEbound0}. Therefore this particular $\var$ is achievable with $T$ resources. But we can do better. The value $x_K$ obtained from Eq.~\eqref{eq:totalResource} gives the actual distribution ${\nu_j}=\round{x_j}$ from Eq.~\eqref{eq:linearRamp}. The amount of resources $N$ used in the strategy identified by this specific $x_K$ is given by Eq.~\eqref{nconstraint}. By construction $N \le T$ and we define
\begin{equation}
	\Delta N := T - N = T - 2 \sum_{j=1}^K \round{x_j} 2^{j-1} \, \le 2 \cdot 2^{K} + \mathcal{O} \left( K \right).
\end{equation}
The extra terms $\mathcal{O} \left( K \right)$ arise because of the approximations in Eq.~\eqref{eq:totalResource}. To avoid them we must solve $N_K^> - 2 \gamma \left( K+1 \right) - 2 x_K - 1 = T$ to find $x_K$, instead of $N_K^> = T$. In this section we see how to employ the extra resources $\Delta N$ in order to do slightly better than bound Eq.~\eqref{eq:continousMSEbound} We modify the resource distribution as $x_j = \gamma \left( K - j \right) + x_K + \Delta \nu_j$ with $\Delta \nu_j \in \mathbb{N}$ such that $2 \sum_{j = 1}^{K} \Delta \nu_j 2^{j-1} = \Delta N$. The objective is to optimize on $\Delta \nu_j$ subject to the constraints
\begin{eqnarray}
	\Delta \nu_j > - \round{ \gamma \left( K - j \right) + x_K } \;,
	\label{NEWCONSTRA}
\end{eqnarray}
(so that we don't erase any step of the estimation). Then we rewrite Eq.~\eqref{eq:total} as
\begin{equation}
	\var \le \left( \frac{2 \pi}{3} \right)^2 \left( \frac{1}{4^K} +\frac{64 A }{2^{3 K} C^{x_K - \frac{1}{2}}} \sum_{j=1}^{K} 2^{j - \log_2 C \cdot \Delta \nu_j} \right) \;,
	\label{eq:modifiedSummation}
\end{equation}
where we just accounted for the effect of having the extra measurements at disposal. We see that in order to minimize the MSE we need to minimize the summation $ \sum_{j = 1}^{K} 2^{j - \log_2 C \cdot \Delta \nu_j}$. We will forget about the constraints (\ref{NEWCONSTRA}) as we check in retrospect that our solution satisfies them anyway.
\begin{thm}
	Given the number of additional probes $\Delta N = 2 \sum_{j = 1}^{K} b_j 2^{j-1}$ written in binary representation, the optimal $\Delta \nu_j$ is $\Delta \nu_j = b_j$.
	\label{thm:optimalDeltaM}
\end{thm}
The proof of this theorem is given in Appendix~\ref{app:proof}. It means that we should build with the extra resources states that are as entangled as possible. We compute the MSE bound given by such optimal distribution by using $\sum_{j = 1}^{K} 2^{j- b_j \log_2 C } = \sum_{j = 1}^{K} \left[ 2^j - \left( 1 - \frac{1}{C} \right) b_j 2^{j} \right] = \sum_{j=1}^{K} 2^{j} - \left( 1 - \frac{1}{C} \right) \Delta N$, and it reads
\begin{eqnarray}
	\begin{aligned}
	\var &\le \left( \frac{2 \pi}{3} \right)^2 \left( \frac{1}{4^K} \! + \! \frac{64 A}{2^{3 K} C^{x_K - \frac{1}{2}}} \sum_{j=1}^{K} 2^{j- \log_2 C \cdot b_j} \right) \\ &= \left( \frac{2 \pi}{3} \right)^2 \left( 1 + \frac{128 A}{C^{x_K-\frac{1}{2}}} \right) \frac{1}{4^K} \\ & \quad - \left( \frac{2 \pi}{3} \right)^2 \left(1 - \frac{1}{C} \right) \frac{64 A} {2^{3 K} C^{x_K - \frac{1}{2}}} \Delta N \;,
	\end{aligned}
	\label{eq:optimalRMSEK}
\end{eqnarray}
where $0 \le \Delta N \le 2 \cdot 2^K$. Notice that this formula apparently works only for $\Delta N$ even, as prescribed by Theorem~\ref{thm:optimalDeltaM}, but we consider it valid for every $\Delta N$ (also odd) between $0$ and $2 \cdot 2^K$. For more details see Appendix~\ref{app:smallSteps}. In conclusion we compute an upper bound $\text{QFI}^{>}$ on the QFI of the complete input state in Eq.~\eqref{eq:wholeInput}, modified with $\Delta \nu_j$, starting from Eq.~\eqref{eq:fisherPartial}.
\begin{eqnarray}
	\text{QFI}^> &:=& 2 \sum_{j = 1}^{K} x_j \left( 2^{j-1} \right)^2 \\ &=& \left( \frac{2 \gamma}{3} + 2 x_K + 1 \right) \frac{4^K}{3} + 2 \sum_{j = 1}^{K} 4^{j-1} \Delta \nu_j \;.
\end{eqnarray}
Given that $\frac{\Delta N}{2} = \sum_{j=1}^{K} \Delta \nu_j 2^{j-1} = \sum_{j=1}^{K} b_j 2^{j-1}$ we ask how the extra term $2 \sum_{j = 1}^{K} 4^{j-1} b_j$ compares with $\frac{\Delta N^2}{2} = 2 \left( \sum_{j=1}^{K} b_j 2^{j-1} \right)^2$. If only one $b_j = 1 $ then
\begin{equation}
	2 \sum_{j = 1}^{K} 4^{j-1} b_j = \frac{\Delta N^2 }{2} \;.
\end{equation}
The other extremal case happens when $b_j = 1$ for all $j$, then
\begin{eqnarray}
	2 \sum_{j = 1}^{K} 4^{j-1} b_j &=& \frac{2}{3} \left( 4^K - 1 \right) \\ &\ge& \frac{\Delta N ^2 }{6} = \frac{2}{3} \left( 2^K - 1 \right)^2 \;.
\end{eqnarray}
In general for whatever $b_j$ it holds
\begin{equation}
	\frac{\Delta N^2}{6} \le 2 \sum_{j = 1}^{K} 4^{j-1} b_j \le \frac{\Delta N^2}{2} \;.
\end{equation}
Therefore we have the following two bound for $\text{QFI}^{>}$, i.e. 
\begin{eqnarray}
	 && \text{QFI}^{>} \ge \left[ \left( \frac{2 \gamma}{3} + 2 x_K + 1 \right) \frac{4^K}{3} + \frac{\Delta N^2}{2} \right]^{-1}, \label{eq:lowerBoundQFI} \\ && \text{QFI}^{>} \le \left[\left( \frac{2 \gamma}{3} + 2 x_K + 1 \right) \frac{4^K}{3} + \frac{\Delta N^2}{6} \right]^{-1}. \label{eq:upperBoundQFI}
\end{eqnarray}

\section{Optimal distribution in the presence of external limitations}
\label{sect:external}
In this section we study two situations where some external constraints affect the the estimation process limiting its precision and forcing us to modify the optimal strategy. The first one is the case in which the maximum allowed dimension of the entangled state is limited (by technological constraint for example) and its much smaller than the entangled size required for the optimum strategy with a given number of resources $N$. In such case when $N \rightarrow \infty$ the precision of the estimation with the resource distribution of section Sec.~\ref{sect:algorithm} is not optimal. Then an hybrid strategy, which explicitly consider an estimation at the SQL in the last step will be a better choice. The second scenario consists in the addition of a loss noise. We will compare the optimal distributions of an amount of resources $N$ in the noisy and noiseless case, both without further constraints and with a maximum entanglement size constraint.

\subsection{Optimal distribution of resources with limited entanglement}
\label{subsect:limitedEnt}
Consider the case where we are allowed to entangled our states only up to a size $2^{K-1}$, for some given integer value $K$. Under this circumstance the possibility of reaching HS~(\ref{HSnew}) in the large $N$ limit, is clearly prevented as one can easily verify by looking at the inequality~(\ref{eq:explicitQCRbound}). Yet we may consider the possibility of using an hybrid strategy that employs the entanglement resources we are provided to reach a $1/N$ SQL for the MSE with an optimal factor. The idea is to use maximally entangled states of sizes $M_1=1,M_2= 2, M_3=4, \dots,M_{K-2}= 2^{K-2}$ to progressively restrict the search region, and then employ $2 \nu_K \gg 1$ copies of a GHZ-like state of maximal size $M_{K-1}=2^{K-1}$ to produce an estimator $\hat{\theta}_K$ that saturates the QCR bound \eqref{eq:explicitQCRbound}, i.e.
\begin{equation}
	\Delta^2 \hat{\theta}_{K} = \int |\hat{\theta}_{K} - \theta|^2 P (\hat{\theta} | \theta) = \left( \frac{1}{2^{K-1}} \right)^2 \frac{1}{2 
	\nu_K} \;,
	\label{eq:lastEstimator}
\end{equation}
a possibility that is e.g. granted by using the adaptive measurement discussed in Ref.~\cite{Fujiwara2006} -- see Appendix~\ref{app:adaptive} for details. In order to determine the optimal choice of the parameters $\nu_j$, we can use the bound~\eqref{eq:RMSEBound} where now we substitute the last precision range (reached if all the previous steps were correct) with the MSE in Eq.~\eqref{eq:lastEstimator}. The solution can hence be founded by studying the associated Lagrangian problem 
\begin{multline}
	\mathcal{L} := \left( \frac{1}{2^{K-1}} \right)^2 \frac{1}{2 x_K} + \sum_{j = 1}^{K-1} \left( \frac{8 \pi}{3 \cdot 2^{j-1}} \right)^2 A C^{-x_j} \\ - \lambda \left(2 \sum_{j = 1}^{K} 2^{j-1} x_j - N \right) \;,
	\label{eq:entLimLagrangian}
\end{multline}
where, as in the case detailed in Sec.~\ref{subsect:proofHeisenberg}, we treat the integer variables $\nu_j$ as real quantities $x_j$. The derivatives with respect to $x_j$ read
\begin{align}
	&\partial_{x_K} \mathcal{L} = - \left( \frac{1}{2^{K-1}} \right)^2 \frac{1}{2 x_K^2} - \lambda 2^K = 0 \;, \\ &\partial_{x_j} \mathcal{L} = - \left( \frac{2 \pi}{3} \right)^2 \frac{16 A \log C}{4^{j-1}} C^{-x_j} - \lambda 2^{j} = 0 \;,
\end{align}
where the first one holds for $j = K$ and the second is for $j \le K-1$. Having obtained $\lambda$ from the $(j-1)$-th derivative we compute $x_K$ as a function of $x_{K-1}$, obtaining
\begin{equation}
	x_{K} = \frac{3 C^{\frac{x_{K-1}}{2}}}{2 \pi \left( 256 A \log C \right)^\frac{1}{2}} \;,
	\label{eq:optimalnuK}
\end{equation}
which in order to deliver the value of $\nu_K$ should be rounded to the nearest integer (notice however that since we expect $\nu_{K} \gg 1$ the rounding doesn't play any role in $\var$). The optimal number of measurements performed in the last step (with states of size $2^{K-1}$) grows exponentially in the number of measurements used in the previous localization phase. The other $x_j$ for $j \le K-1$ are 
\begin{equation}
	x_j = \gamma \left( K-1-j \right) + x_{K-1} \, ,
	\label{eq:linearRamp2}
\end{equation}
which again should be rounded to the nearest integer. The localization steps from $j=1$ to $j=K-1$ operate at the Heisenberg scaling but the great majority of the resources is employed in the last step that operates at the Standard Quantum Limit. The resummed upper bound on the MSE is hence 
\begin{multline}
	\var \le \frac{1}{4^{K-1}} \frac{\pi}{3} \left( 256 A \log C \right)^{\frac{1}{2}} C^{-\frac{x_{K-1}}{2}} \\ + \frac{1}{4^{K-1}} \left( \frac{2 \pi}{3} \right)^2 \frac{128 A}{C^{x_{K-1} - \frac{1}{2}}} \;,
	\label{eq:RMSEBoundLimitedEnt}
\end{multline}
while the resource summation equation instead gives
\begin{equation}
	N = 2 \sum_{j=1}^{K-1} \round{x_j} 2^{j-1} + \Big \lfloor \frac{3 C^{\frac{x_{K-1}}{2}}}{2 \pi \left( 256 A \log C \right)^\frac{1}{2} } \Big \rceil 2^K \;.
	\label{eq:totalprobeEntLimited}
\end{equation}
For growing $\nu_{K-1}$ its clear how the MSE is dominated by the first term coming from the SQL.
\begin{figure}[!t]
	\begin{center}
		\includegraphics[width=0.45\textwidth]{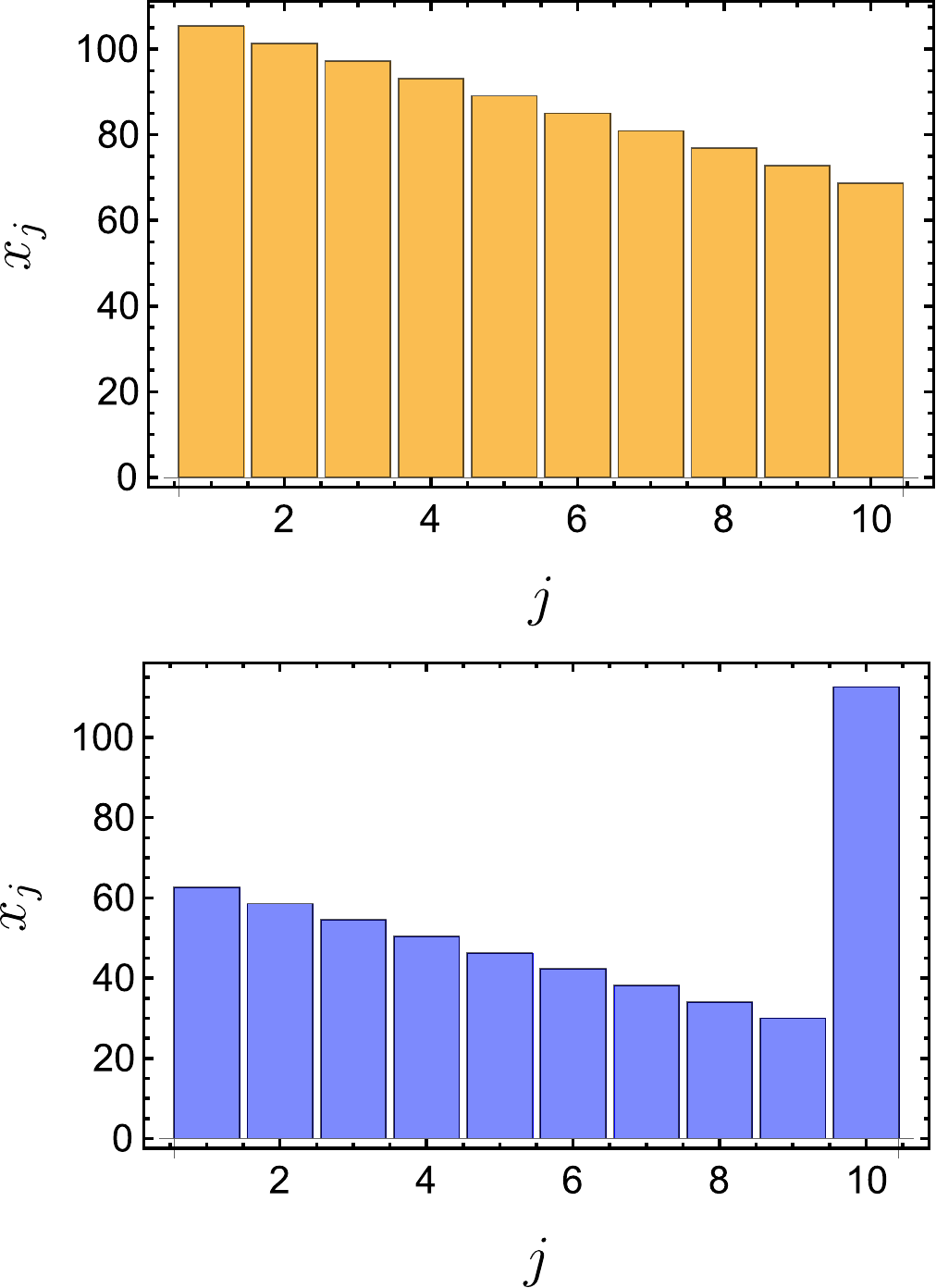}
	\end{center}
	\caption{Both distributions $x_j$ refer to $K = 10$ and to almost the same number of probes $N \simeq 1.5 \times 10^5$. The upper (orange) chart is the linear ramp in Eq.~\eqref{eq:linearRamp} with $x_K = 68.7$, while the lower (blue) chart is the distribution in Eq.~\eqref{eq:linearRamp2} with $x_{K-1} = 30$ for $j = 1, \dots, K-1$ and Eq.~\eqref{eq:optimalnuK} for $j = K$. The upper bound on the MSE of the first distribution saturate to the limit $x_j \rightarrow \infty$, this can be check using the analytical values of $A$ and $C$. The bounds are respectively $\Delta^2 \hat{\theta} \le 4.18 \cdot 10^{-6}$ for Eq.~\eqref{eq:linearRamp} and $\Delta^2 \hat{\theta} \le 1.73 \times 10^{-8}$ for the limited entanglement optimized strategy. }
	\label{fig:chartLimited}
\end{figure}
The derived entanglement limited optimal strategy is compared to that in Eq.~\eqref{eq:linearRamp} in Fig.~\ref{fig:chartLimited}. Here we neglected that $x_j$ are not integers and set $N = 2 \sum_{j=1}^K x_j 2^{j-1}$. The tendency is to reduce the number of resources used for steps $j \le K-1$ and concentrate them to the biggest entangled state constructible. As $N$ grows the MSE approaches the CR bound $2^{K-1}/N$. The previous analysis can be easily generalized to account also for the case where we are bound to use states of size at most $R$ with $R$ being an arbitrary integer not necessarily multiple of 2. then we could employ a series of states of sizes $R, \ceil{\frac{R}{2}}, \ceil{\frac{R}{4}}, \dots, 1$. For the last states, on which effectively depends the MSE, we can neglect the fact that $\frac{R}{2}$, $\frac{R}{4}, \dots$ are not integers (because $R \gg 1$), therefore we write a simple and suitable Lagrangian
\begin{multline}
	\mathcal{L} := \frac{1}{2 R^2 x_K} + \sum_{j=1}^{K-1} \left( \frac{8 \pi \cdot 2^{K-j}}{3 R} \right)^2 A C^{- x_j} \\ -\lambda \left[ 2 \sum_{j=1}^{K} x_j \left( \frac{R}{2^{K-j}} \right) - N \right] \;,
	\label{eq:lagragianR}
\end{multline}
where also for the states with fewer probes we haven't rounded the size as their corresponding terms will not affect much the error. In Eq.~\eqref{eq:lagragianR} $K$ is chosen to be the smallest value for which $\ceil{\frac{R}{2^{K-1}}} = 1$. We notice that the MSE is rescaled by a factor $\left( \frac{2^{K-1}}{R} \right)^2$, while $\frac{R}{2^{K-1}}$ is the rescaling of the total number of probes. Therefore by defining $\kappa := \frac{R}{2^{K-1}}$ we have the Lagrangian
\begin{multline}
	\mathcal{L} := \frac{1}{\kappa^2} \left( \frac{1}{2^{K-1}} \right)^2 \frac{1}{2 x_K} + \frac{1}{\kappa^2} \sum_{j = 1}^{K-1} \left( \frac{8 \pi}{3 \cdot 2^{j-1}} \right)^2 A C^{-x_j} \\ - \lambda \left(2 \kappa \sum_{j = 1}^{K} 2^{j-1} x_j - N \right) \;.
\end{multline}
The optimal $x_j$ are again given by Eq.~\eqref{eq:optimalnuK} and Eq.~\eqref{eq:linearRamp2}, the only difference being that in the resource summation~\eqref{eq:totalprobeEntLimited} we substitute $N \rightarrow \frac{2^{K-1} N}{R}$.

\subsection{Optimal distribution of resources with noise}
\label{subsect:distrNoise}
\begin{figure}[!t]
	\begin{center}
		\includegraphics[width=0.45\textwidth]{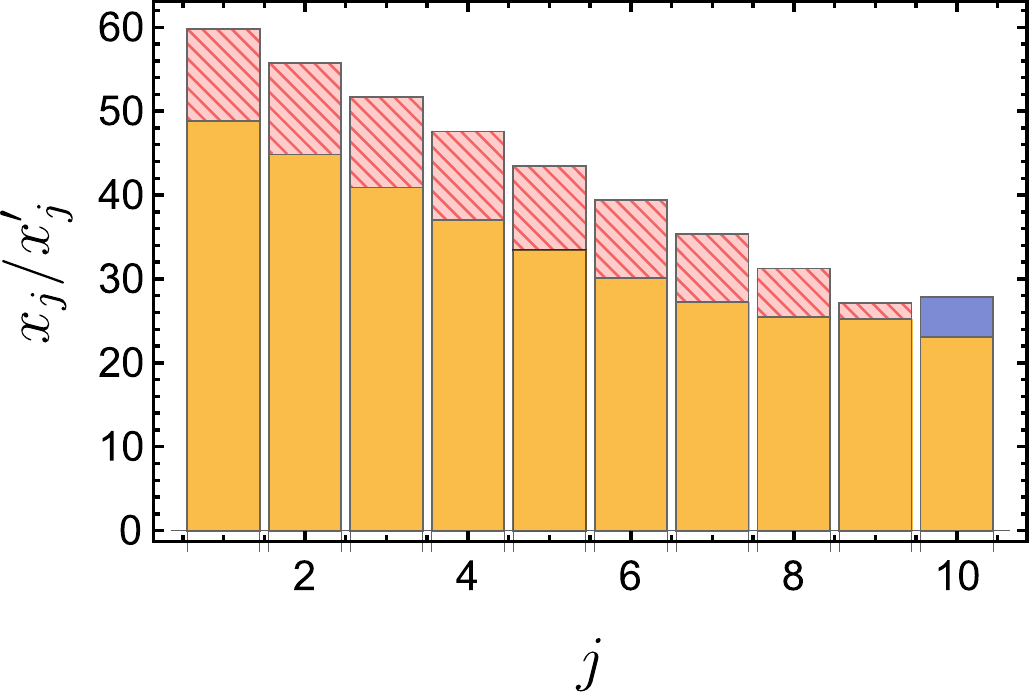}
	\end{center}
	\caption{The solid blue bar and red striped bars are respectively the number of states to be added (solid bar) or subtracted (striped bar) at each level $j$ of the estimation according to Eq.~\eqref{eq:noisynujprimed} with $x_K = 10$ and $\eta = 0.998$, with respect to the base noiseless strategy given in Eq.~\eqref{eq:linearRamp} with $x_K = 23.1$. Both strategies refer approximately to the same number of probes $N \simeq 5.6 \times 10^4$ and to $K = 10$. The numerical values for $A$ and $C$ are those of bound~\eqref{eq:expBoundNumeric}. The number of states to be used in the noisy strategy exceeds that of the noiseless one only in step $j = K$.}
	\label{fig:noisy1}
\end{figure}

\begin{figure}[!t]
	\begin{center}
		\includegraphics[width=0.45\textwidth]{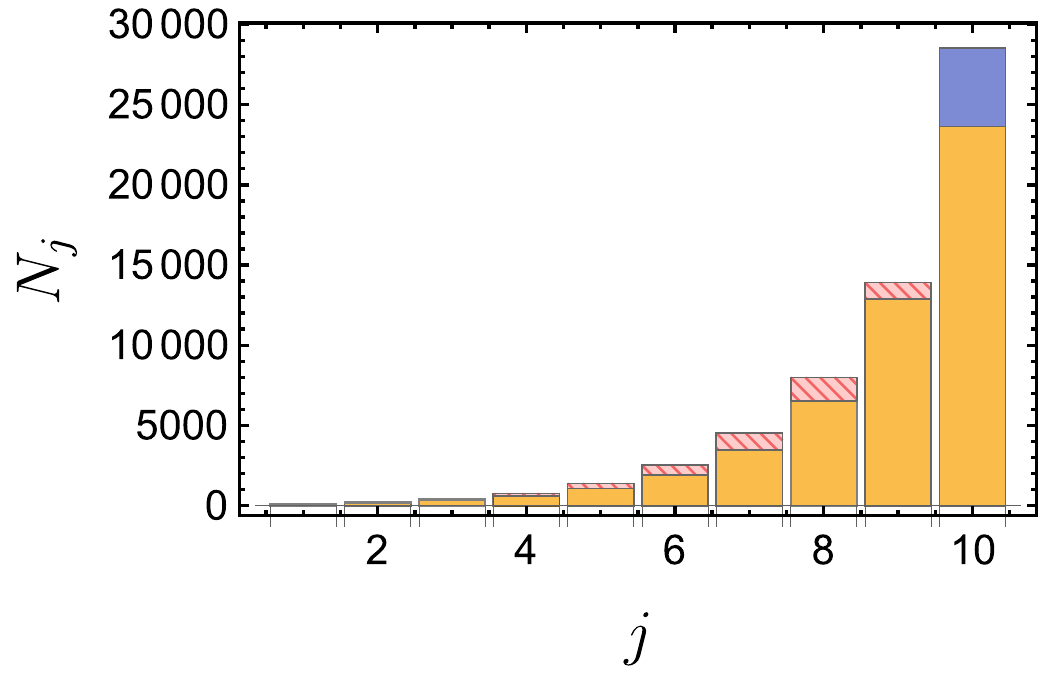}
	\end{center}
	\caption{The solid blue bar and red striped bars are respectively the number of probes to be added (solid bar) or subtracted (striped bars) at each level $j$ of the estimation according to Eq.~\eqref{eq:noisynujprimed} with $x_K = 10$ and $\eta = 0.998$, with respect to the base noiseless strategy given in Eq.~\eqref{eq:linearRamp} with $x_K = 23.1$. Both strategies refer approximately to the same number of probe $N \simeq 5.6 \times 10^4$ and to $K = 10$. The resources of the noisy strategy are $N_j = x'_j 2^{j}$ while that of the base are $N_j = x_j 2^{j}$. The numerical values for $A$ and $C$ are those of bound~\eqref{eq:expBoundNumeric}. Resources are reallocated to the highest entangled states from less entangled regions. Notice that the step which in absolute terms is stripped off more of resources is $j=8$, in relative terms it is $j=6$.}
	\label{fig:noisy2}
\end{figure}

Consider now a simple case in which loss is added to the probes, this will be characterized by the value $\eta$, meaning that there is a probability $\eta$ of retaining a certain probe and $1 - \eta$ of losing it. This is particularly damaging for the maximally entangled states, as a GHZ state of size $2^{j-1}$ can survive only with probability $\eta^{2^{j-1}}$. The expression of the Lagrangian to minimize in this scenario is
\begin{multline}
	\mathcal{L} := \left( \frac{2 \pi}{3} \right)^2 \left( \frac{1}{4^K} + 16 \sum_{j = 1}^{K} \frac{A}{4^{j-1}} C^{-x_j} \right) \\ - \lambda \left( 2 \sum_{j=1}^{K} \frac{x_j}{\eta^{2^{j-1}}} 2^{j-1} - N \right) \;.
	\label{eq:lagrangianNoise}
\end{multline}
The parameter $x_j$ is the number of measurements we expect to perform at step $j$ after the loss, so it appears in the probability of error $A C^{-x_j}$. However the expected number of probes to be employed, accounting also those that will be lost, is $x_j' := \frac{x_j}{\eta^{2^{j-1}}}$, which appears in the constraint of the Lagrangian. These numbers have to be rounded to refer to the actual strategy. The derivative with respect to $x_j$ gives
\begin{equation}
	\lambda = - \left( \frac{2 \pi}{3} \right)^2 \frac{16 A \log C}{2^{3 j - 2}} C^{- x_j} \eta^{2^{j-1}} \;.
\end{equation}
Also in this case the optimal distribution of the resources can be found analytically by imposing the equation
\begin{equation}
	- x_j \log_2 C + \frac{\log \eta}{\log 2} 2^{j-1} - 3 j = \text{const.} \; ,
\end{equation}
which gives the expressions
\begin{equation}
	x_j = \gamma \left( K - j \right) + x_K + \frac{| \log \eta |}{\log C} \left( 2^{K-1} - 2^{j-1} \right) \;,
	\label{eq:noisynuj}
\end{equation}
and
\begin{equation}
	x'_j = \frac{\gamma \left( K - j \right) + x_K}{\eta^{2^{j-1}}} + \frac{| \log \eta |}{\log C} \frac{ 2^{K-1} - 2^{j-1} }{\eta^{2^{j-1}}} \; .
	\label{eq:noisynujprimed}
\end{equation}
A proper comparison between the noisy and the noiseless optimal distributions is to be carried out between strategies referring to the same number of resources $N$, hence having different $x_K$. Such fair comparison is presented in Fig.~\ref{fig:noisy1} and Fig.~\ref{fig:noisy2}, these show the reallocation of the probes in the various steps. We took the number of resources to be $N = 2 \sum_{j=1}^K x_j 2^{j-1}$ for the noiseless strategy and $N = 2 \sum_{j=1}^K x'_j 2^{j-1}$ for the noisy one, avoiding the rounding, as we want to show only the main differences not precise numerical results. The comparison tells us that the resources are expected to migrate toward the high entanglement end, as these are the states more affected by the loss.
As in the precedent subsection we can limit the entanglement size to $R$ and write the following Lagrangian for the resource optimization when noise is present
\begin{multline}
	\mathcal{L} := \frac{1}{\kappa^2} \left( \frac{1}{2^{K-1}} \right)^2 \frac{1}{2 x_K} + \frac{1}{\kappa^2} \sum_{j = 1}^{K-1} \left( \frac{8 \pi}{3 \cdot 2^{j-1}} \right)^2 A C^{-x_j} \\ - \lambda \left(2 \kappa \sum_{j = 1}^{K} 2^{j-1} x_j \eta^{-\kappa 2^{j-1}} - N \right) \;.
\end{multline}
The derivatives with respect to $x_j$ are
\begin{align}
	& \partial_{x_K} \mathcal{L} = -\frac{1}{4^{K-1}} \frac{1}{2 \kappa ^2 x_K^2} - \kappa \lambda 2^K \eta^{-\kappa 2^{K-1}} = 0 \;, \\ &\partial_{x_j} \mathcal{L} = - \left( \frac{2 \pi}{3} \right)^2 \frac{16 A \log C}{\kappa^2 4^{j-1}} C^{-x_j} - \kappa \lambda 2^{j} \eta^{-\kappa 2^{j-1}} = 0 \;,
\end{align}
they give $x_K$ as function of $x_{K-1}$, i.e.
\begin{equation}
	x_{K} = \frac{3 \eta^{\frac{R}{4} }C^{\frac{x_{K-1}}{2}}}{2 \pi \left( 256 A \log C \right)^\frac{1}{2}} \;.
	\label{eq:noisynuk}
\end{equation}
When $N \rightarrow \infty$ we have $N \simeq \frac{2 R}{\eta^R} x_K$ and $\var \simeq \frac{1}{2 R^2 x_K} = \frac{1}{R \eta^R N}$, which is exactly the inverse of Eq.~\eqref{eq:fisherTrunc}. Therefore given a certain level of noise we can choose the optimal maximum size of the entangled states (see Sec.~\ref{subsect:probeLoss}) and obtain an asymptotic SQL scaling with prefactor which is the best allowed for a GHZ-like state, all after a localization procedure at the Heisenberg scaling.

\subsection{The GHZ state in presence of loss}
\label{subsect:probeLoss}
In the presence of loss the entanglement size is naturally limited to those states that are metrologically useful. Indeed the QFI for a GHZ state of size $N$ subject to a loss $\eta$ is
\begin{equation}
	\text{QFI} \left( |{\rm GHZ} ^{(N)}\rangle \right) = \eta^N N^2 \;.
\end{equation}
\begin{figure}[!t]
	\begin{center}
		\includegraphics[width=0.45\textwidth]{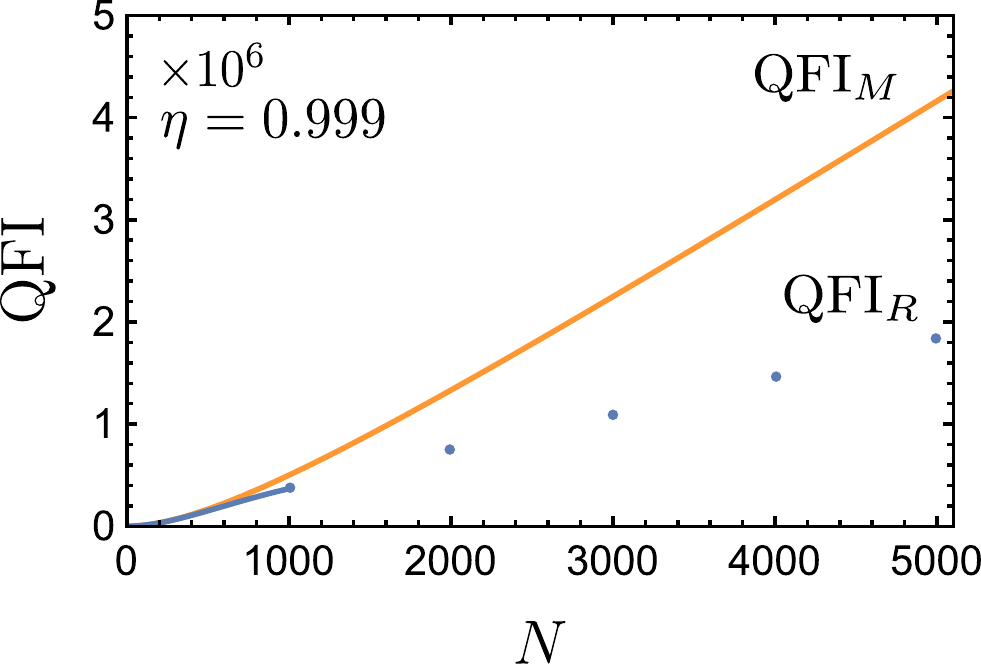}
	\end{center}
	\caption{Comparison between $\text{QFI}_M$ and $\text{QFI}_R$, the latter being discrete as the entangled states have size $R$.}
	\label{fig:scalingSQL}
\end{figure}
This drops quickly to zero after a maximum size dependent on $\eta$. This type of noise is the qubit equivalent of photon loss in both arms of an interferometer. Given $N$ resources, they can be divided in bunches of $R$ probes to be entangled~\cite{Dorner2009}, so that the asymptotic QFI of the process will scale linearly as
\begin{equation}
	\text{QFI}_R := \eta^{R} R^2 \; \frac{N}{R} \;.
	\label{eq:fisherTrunc}
\end{equation}
By maximizing this expression we find the optimal cut $R = -\frac{1}{\log \eta}$, which corresponds to $\text{QFI}_R := - \frac{N}{e \log \eta}$. We compare this with an upper bound valid for every state~\cite{Escher2011, Kolodynski2015}, when noise is present, i.e.
\begin{equation}
	\text{QFI} \le \text{QFI}_{M} := \frac{N^2}{1+\frac{1-\eta}{\eta} N} \;,
\end{equation}
see Fig.~\ref{fig:scalingSQL}. The asymptotic ratio between the upper bound QFI$_M$ and the one obtained by employing suitable GHZ-like states is
\begin{equation}
	\kappa := \lim_{N \rightarrow +\infty} \frac{\text{QFI}_M}{\text{QFI}_R} = -\frac{e \eta \log \eta }{1 - \eta} \;,
\end{equation}
so we see that the precision bound using only GHZ-like states is at most a factor $\sim \sqrt{e} = 1.65$ away from that of the absolute optimal state. The state size $R$ can be reached at the end of a procedure of localization employing smaller states, like the one presented in the paper. The probability of not being in the correct window drops exponentially and the MSE asymptotically scales as $\var \sim -\frac{e \log \eta}{N}$.

\section{Conclusions}
\label{CONCSEC} 
Quantum metrology has shown how it is possible to exploit the hypersensitivity of the entangled states to boost the phase estimation task. A particular attention is dedicated in the literature to maximally entangled states. It is though often unspoken that because these states cycle multiple times when subject to phase encoding they erase the large scale information on the position of the phase, while holding the information regarding its small fluctuation. This appears to make maximally entangled states unsuitable to be used alone without a prior localization of the phase, see Sec.~\ref{sect:phaseEstimation}. The risk is that because of the necessity of this preliminary stage, with its resource requirement, the Heisenberg scaling is lost. In this paper we studied the scenario in which such localization is performed by maximally entangled states growing exponentially in size, which step by step codify finer and finer properties of the unknown phase. We reviewed here that by suitably choosing the number of states of each size it is possible to prove rigorously that HS is still achieved, see Sec.~\ref{subsect:proofHeisenberg}. A proper distribution of resources is what above all allows for the HS. The actual physical operations to be performed in an experiment are pretty straightforward. Given the total amount of probes one has to create entanglement in it according to the distribution in Eq.~\eqref{eq:linearRamp} and to Sec.~\ref{secB} and Sec.~\ref{subsect:interpolation}, then he can encode and measure each individual probe. There is no need of performing whatever type of entangled measurement as single probe measurement will suffice if they are distinguishable, see Sec.~\ref{subsect:build} and Appendix~\ref{app:SEPMES}. As a matter of fact also entanglement among the probes can be exchange for multipassage through the phase encoding process. The measurements output can then be saved and later processed to find the estimator. The data analysis stands as a simple task, however we reported explicitly (for the sake of completeness) a valid pseudocode in Algorithm~\ref{alg:nonAdaptive}.
Because the procedure is non-adaptive the measurement stage and the data processing stage are completely independent. If further measurement are conducted then they can just be added to the data set which will be reprocessed.

The optimization results are relative to the upper bound~\eqref{eq:RMSEBound} on the MSE, and are definitive to this regard. Of course such inequality does not necessarily predict the actual results of the algorithm (which could be better than the one dictated by the bound): nevertheless this is the furthest we could carry out an analytical approach. A further improvement toward obtaining the optimal resource scheme for the algorithm could come from tighter bounds or numerical computations, but we decided purposefully to avoid as much as possible numerics in order to give an analytic review. The analysis of the limited entanglement and noisy cases are to be thought more as toy models, still they capture some of the key features of those scenarios.

\section*{ACKNOWLEDGMENTS}
We thank Shelby Kimmel and Howard M. Wiseman for their feedbacks.
We acknowledges support by MIUR via PRIN 2017 (Progetto di Ricerca di Interesse Nazionale): project QUSHIP (2017SRNBRK).

\appendix

\section{Separability of the optimal projective measurement} 
\label{app:SEPMES} 
Here we explicitly show that the Type-$0$ measurement (as well as the Type-$+$) can obtained via a separable procedure~\cite{Bollinger1996}. As for any probe only two quantum states are involved in the construction of the GHZ-like state, from now on we simply assume they are provided by qubit systems and use the associated standard notation. Given hence the output state~(\ref{OUTPUTPHI}) we observe that by applying an Hadamard gate to each of the probes that compose it, we can transform it into the following vector 
\begin{multline}
	\frac{1}{\sqrt{2}} \left[ \left( \frac{\ket{0} + \ket{1}}{\sqrt{2}} \right)^{\otimes M} + e^{i M \theta } \left( \frac{\ket{0} - \ket{1}}{\sqrt{2}} \right)^{\otimes M} \right] \\ = \frac{1}{2^{\frac{M+1}{2}}} \sum_{k =0}^{M} \sqrt{\binom{M}{k}} \ket{M-k, k} \left[ 1 +e^{i M \theta} \left( -1 \right)^k \right] \;, \label{nuovostato} 
\end{multline}
where for easy of notation we replace $M_j$ with $M$. In the second line of the above expression $\ket{M-k, k}$ is a normalized and symmetrized state and corresponds to $M-k$ probes in the state $\ket{0}$ and $k$ in $\ket{1}$. We then project each probe of the transformed state~(\ref{nuovostato}) on their corresponding computational basis. The probability of getting $k$ probes in the state $\ket{1}$ is 
hence given by 
\begin{equation}
	p_k = \frac{1}{2^{M}} {\binom{M}{k}} \left[ 1 + \left( -1 \right)^k \cos M \theta \right] \;.
\end{equation}
The phase $M \theta $ modulates the probability outline for odd and even $k$ in the same way, and all the information about $\theta$ is contained in the parity of the probe number. Interestingly enough the probability of getting an odd count is exactly coincident with the probability $p_0$ reported in Eq.~(\ref{eq:adaptiveEstimatorProbability}), i.e. 
\begin{equation}
	 \sum_{k \, \text{odd}} \frac{1}{2^M} \binom{M}{k} \left( 1 + \cos M \theta \right) = \frac{1+\cos M \theta }{2}=p_0 \;.
\end{equation}
This shows that a simple data-processing of the outcomes obtained by the separable measurement detailed above exactly matches the statistical properties of the Type-$0$ detection reported in the main text. Similar conclusions can also be drawn for the Type-$+$ measurement setting: indeed this last can just be obtained from Type-$0$ by adding a proper $\pi/2$ phase shift on the input state, via the action of $V_{\phi} := e^{-i \phi H}$, see App.~\ref{app:adaptive}.

It is worth observing that the possibility of turning Type-$0$ and Type-$+$ measurements into the separable detections schemes, strongly relays on the distinguishability character of the employed probes (a feature that is built-in the qubit model). This property will not be applicable for instance if we consider an estimation task that involves a phase $\theta$ codified in one of the two arms of a Mach-Zehnder interferometer~\cite{SCIARRINO2020, Dowling2015}, with the objective of estimating it through the injection of a limited number of photons detected after the closing beam splitter. Given $a$ and $b$, being the two spatial modes corresponding to the upper and lower arms, the encoding of the phase $\theta$ is performed by a unitary $U_\theta = e^{i \theta N_a}$, where $N_a = a a^\dagger$. It's well known that a path-entangled N00N state is $N$ times more sensitive to the unknown phase than a single photon state~\cite{Giovannetti2011}, indeed
\begin{equation}
	\ket{\text{NOON}_\theta} = (\ket{N0} + e^{i N \theta} \ket{0N}) / \sqrt{2} \;,
\end{equation}
with its QFI being $N^2$, plays the same role of 
the GHZ-like states we consider in the main text. Given that, the protocol discussed in this paper can also be employed in the optical case with the only difference that the necessary photon parity measurements~\cite{Chiruvelli2009} will not be implementable via a separable scheme. Of course this distinction does not apply if each photon is loaded in a different interferometer, each with its own version of the black box $U_\theta$, all identical, then the photons are distinguishable and the optimal measurement is again separable and can be realized with photon counting. 

\section{Derivation of the condition~(\ref{eq:widehatEstimator})}
\label{EQUIVALENCE}
\begin{figure}[!t]
	\begin{center}
		\includegraphics[width=0.48\textwidth]{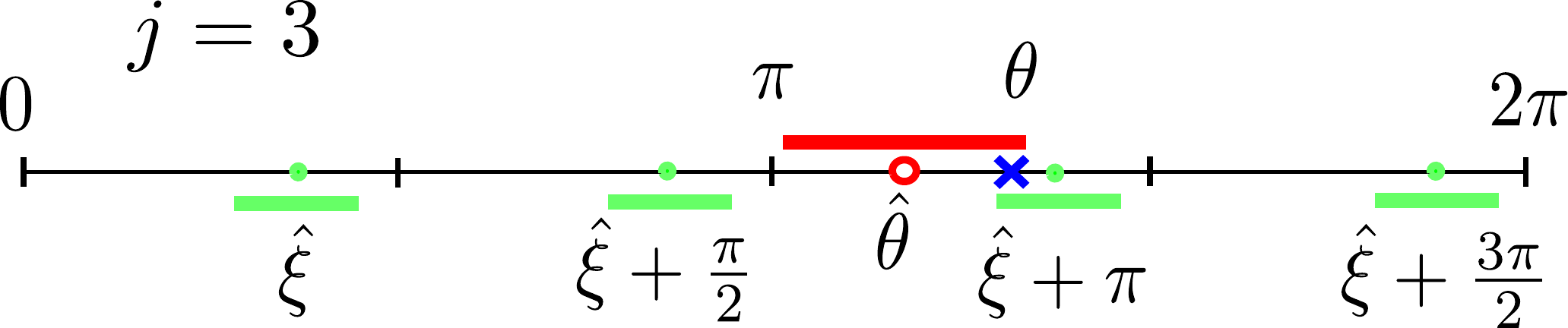}
	\end{center}
	\caption{The hollow red point is the current estimator $\hat{\theta}$ with its confidence interval in red (dark gray), while all the shifted positions of $\hat{\xi}$ (defined in line 5) are in green (light gray). The blue cross is the true value of the parameter $\theta$. Only one of the green (light gray) intervals intersects the red (dark gray) one. The picture refers to $j = 3$ and $M_3 = 4$.}
	\label{fig:line}
\end{figure}

Here we explicitly show that imposing 
\begin{eqnarray} 
	| \hat{\theta}- \theta | \le \frac{\pi}{ 3 \cdot 2^{j-1}} \;,
	\label{eq:condition}
\end{eqnarray}
for all $j$, is equivalent to assume~(\ref{eq:widehatEstimator}). For this purpose it is worth to take a closer look at the various steps of Algorithm~\ref{alg:nonAdaptive}. First of all, in line 1, the estimator is initialized to zero, then in line 3 the size of the $j$-th entangled state is set to $2^{j-1}$ and after measuring its imprinted counterpart the estimator~\eqref{eq:estimator} is computed. In line 5 the variable $\hat{\xi}$ is loaded with $\widehat{M_j \theta} / M_j$. In order to understand line 6 it helps looking at Fig.~\ref{fig:line}. For each $j$ in the cycle we assume that the preceding step (i.e. the $j-1$-th step) of the algorithm was successful so that we can guarantee that given $\hat{\theta}$ the estimator of $\theta$ we have constructed at this point of the procedure, we have
\begin{eqnarray} 
	| \hat{\theta} - \theta | \le \frac{\pi}{ 3 \cdot 2^{j-2}}\;,
	\label{condj-1}
\end{eqnarray}
where as mentioned in the introduction, due to the periodicity of the angular variable, the left-hand-side term is thought to be computed on the unit circle, see Fig.~\ref{fig:unitcircle}. Given the partition $\left[ k \frac{\pi}{2^{j-2}}, k \frac{\pi}{2^{j-2}} + \frac{\pi}{2^{j-2}} \right)$ for $k = 0$ to $k = 2^{j-1} -1$ of $\left[ 0, 2 \pi \right)$, we want to find the one extremum of this partition closest from below to the interval identified by Eq.~(\ref{condj-1}) in which by hypothesis lays the true value of the phase $\theta$. In order to do so we compute
\begin{equation}
	m := \Big \lfloor \frac{\hat{\theta} - \frac{\pi}{3 \cdot 2^{j-2}}}{\frac{\pi}{2^{j-2}}} \Big \rfloor = \Big \lfloor \frac{2^{j-2} \hat{\theta}}{\pi} - \frac{1}{3} \Big \rfloor \;.
	\label{defm} 
\end{equation}
By shifting $\hat{\xi}$ of $\frac{m\pi}{2^{j-2}}$ (line 7) we get near to the previous assessed interval around $\hat{\theta}$. The possible new positions for $\hat{\theta}$ are $\hat{\xi} - \frac{\pi}{2^{j-2}}$, $\hat{\xi}$ and $\hat{\xi}+\frac{\pi}{2^{j-2}}$. By geometric reasoning, because of the choice of $m$, one and only one of the three intervals centered in these new possible positions must overlay with the old interval around $\hat{\theta}$. The two conditions for an interval of size $\frac{2 \pi}{3 \cdot 2^{j-1}}$ centered around $\hat{\xi} - \frac{\pi}{2^{j-2}}$ to overlap with $\left[ \hat{\theta} - \frac{\pi}{3 \cdot 2^{j-2}}, \hat{\theta} + \frac{\pi}{3 \cdot 2^{j-2}} \right)$ are
\begin{align}
	& \hat{\xi} - \frac{\pi}{2^{i-2}} + \frac{1}{6} \frac{\pi}{2^{i-2}} \ge \hat{\theta} - \frac{1}{3} \frac{\pi}{2^{i-2}} \;, \\ & \hat{\xi} - \frac{\pi}{2^{i-2}} - \frac{1}{6} \frac{\pi}{2^{i-2}} < \hat{\theta} + \frac{1}{3} \frac{\pi}{2^{i-2}} \;,
\end{align}
and give the condition in line 8 of the algorithm. For the interval around $\hat{\xi} + \frac{\pi}{2^{i-2}}$ the conditions are instead
\begin{align}	
	& \hat{\xi} + \frac{\pi}{2^{i-2}} + \frac{1}{6} \frac{\pi}{2^{i-2}} \ge \hat{\theta} - \frac{1}{3} \frac{\pi}{2^{i-2}} \;, \\ & \hat{\xi} + \frac{\pi}{2^{i-2}} - \frac{1}{6} \frac{\pi}{2^{i-2}} < \hat{\theta} + \frac{1}{3} \frac{\pi}{2^{i-2}} \;,
\end{align}
and become line 10. If neither $\hat{\xi} - \frac{\pi}{2^{j-2}}$ nor $\hat{\xi} + \frac{\pi}{2^{j-2}}$ get to be chosen as estimator then $\hat{\xi}$ is chosen (line 13). In the end (line 15) the estimator $\hat{\theta}$ is casted into $\left[ 0, 2 \pi \right)$. 

Given all these, let's now show the equivalence between (\ref{eq:widehatEstimator}) and (\ref{eq:condition}). To begin with given $m$ as in (\ref{defm}) and noticing that at the end of the $j$-th step $\hat{\theta}$ is obtained by properly shifting $\frac{\widehat{M_j \theta}}{M_j}$, from (\ref{eq:condition}) we can write 
\begin{align}
	\begin{aligned}
		&\Big | \frac{\widehat{M_j \theta}}{M_j} + m \frac{\pi}{2^{j-2}} \left( \pm \frac{\pi}{2^{j-2}} \right) - \theta \Big | \le \frac{\pi}{3 \cdot 2^{j-1}} \Longrightarrow \\ & | \widehat{M_j \theta} + 2 \pi m \left( \pm 2 \pi \right) - M_j \theta | \le \frac{\pi}{3} \Longrightarrow \\ &| \widehat{M_j \theta} - M_j \theta | \le \frac{\pi}{3} \;.
	\end{aligned}
\end{align}
\begin{figure}[!t]
	\begin{center}
		\includegraphics[width=0.4\textwidth]{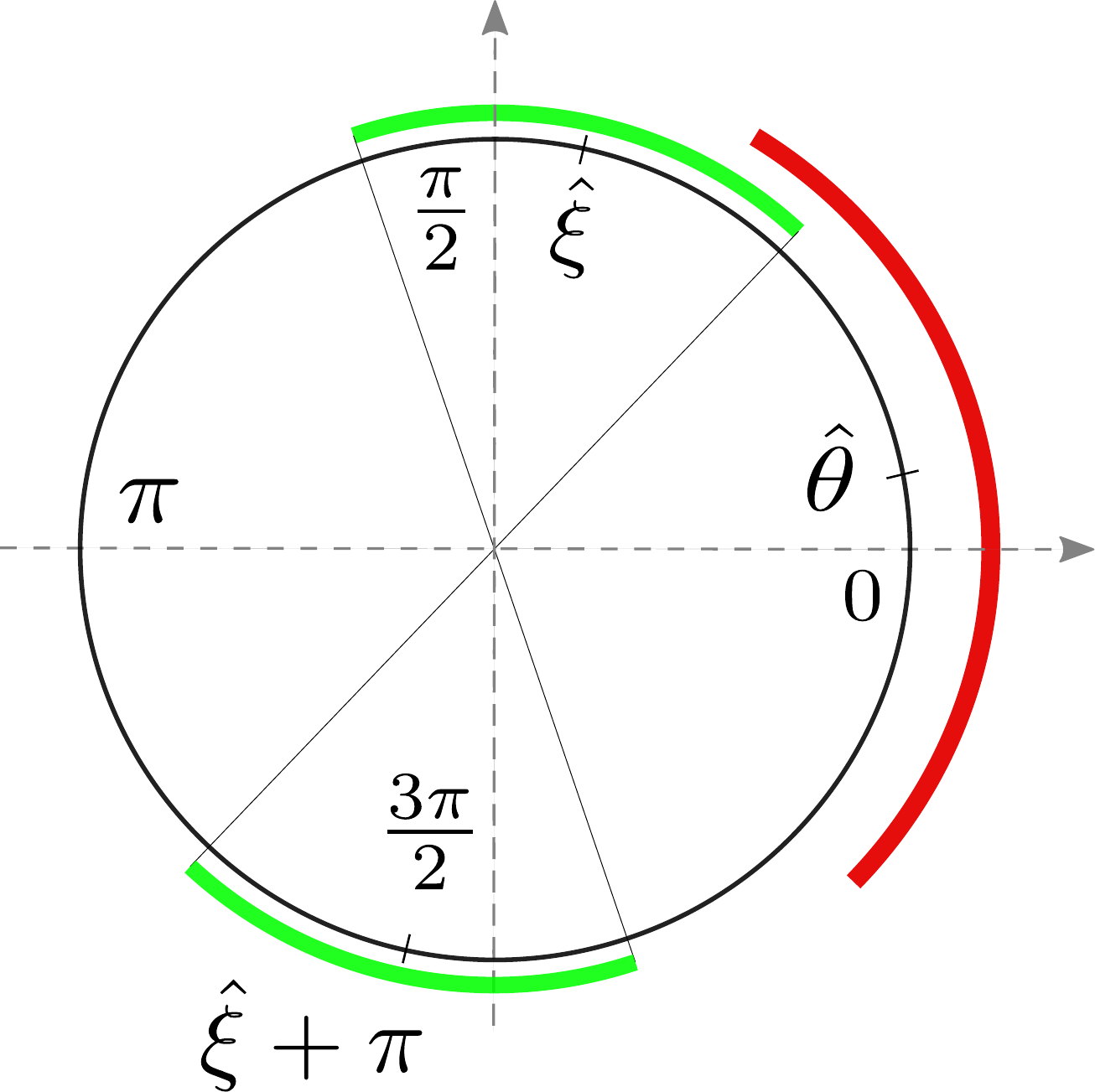}
	\end{center}
	\caption{In this picture we see $\hat{\xi}$, $\hat{\xi}+\pi$, and $\hat{\theta}$ for $j = 2$.}
	\label{fig:circle}
\end{figure}
On the other hand, if all the previous range guess were correct, it is easy to see that the reverse implication holds, see also Fig.~\ref{fig:line}. Indeed given $\hat{\xi} = \widehat{M_j \theta}/M_j$, the condition~\eqref{eq:widehatEstimator} implies $\theta$ to be in one of the intervals
\begin{equation}
	\Big | \hat{\xi} + \frac{k\pi}{2^{j-2}} - \theta \Big | \le \frac{\pi}{3 \cdot 2^{j-1}} \, ,
\end{equation}
with $k = 0, 1, \dots , M_j -1$, these are represented in green (light gray) in Fig.~\ref{fig:line}. Algorithm~\ref{alg:nonAdaptive} selects as $\hat{\theta}$ the only one shifted version of $\hat{\xi}$ which range intersects with the previous known interval for $\theta$, so the range of size $\frac{\pi}{3 \cdot 2^{j-1}}$ centered on the new $\hat{\theta}$ necessarily contains $\theta$, this is expressed by the inequality~\eqref{eq:condition}. 

\section{Alternative choices for the entanglement size parameters $M_j$}
\label{app:genericb}
\begin{figure}[!t]
	\begin{center}
		\includegraphics[width=0.5\textwidth]{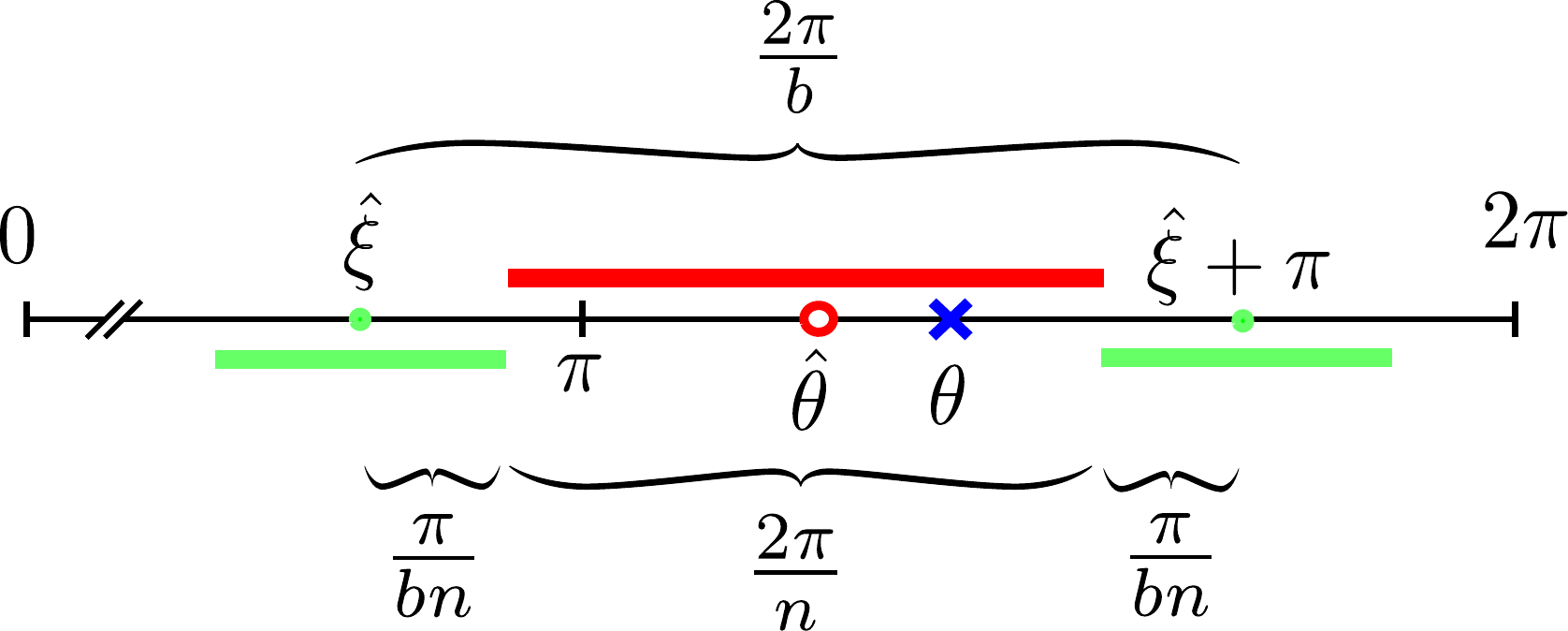}
	\end{center}
	\caption{The picture refers for clarity to $n = 3$ and $b=2$ but the principle is general. The two possible estimators $\hat{\xi}$ and $\hat{\xi} + \pi$ are $\frac{2 \pi}{b}$ apart and their confidence intervals are of size $\frac{2 \pi}{n b}$. In the space between them fits perfectly the previous confidence interval for $\hat{\theta}$, which is of size $\frac{2 \pi}{n}$, therefore we have Eq.~\eqref{eq:conditionIntersection}.}
	\label{fig:line2}
\end{figure}

\begin{figure}[!t]
	\begin{center}
		\includegraphics[width=0.5\textwidth]{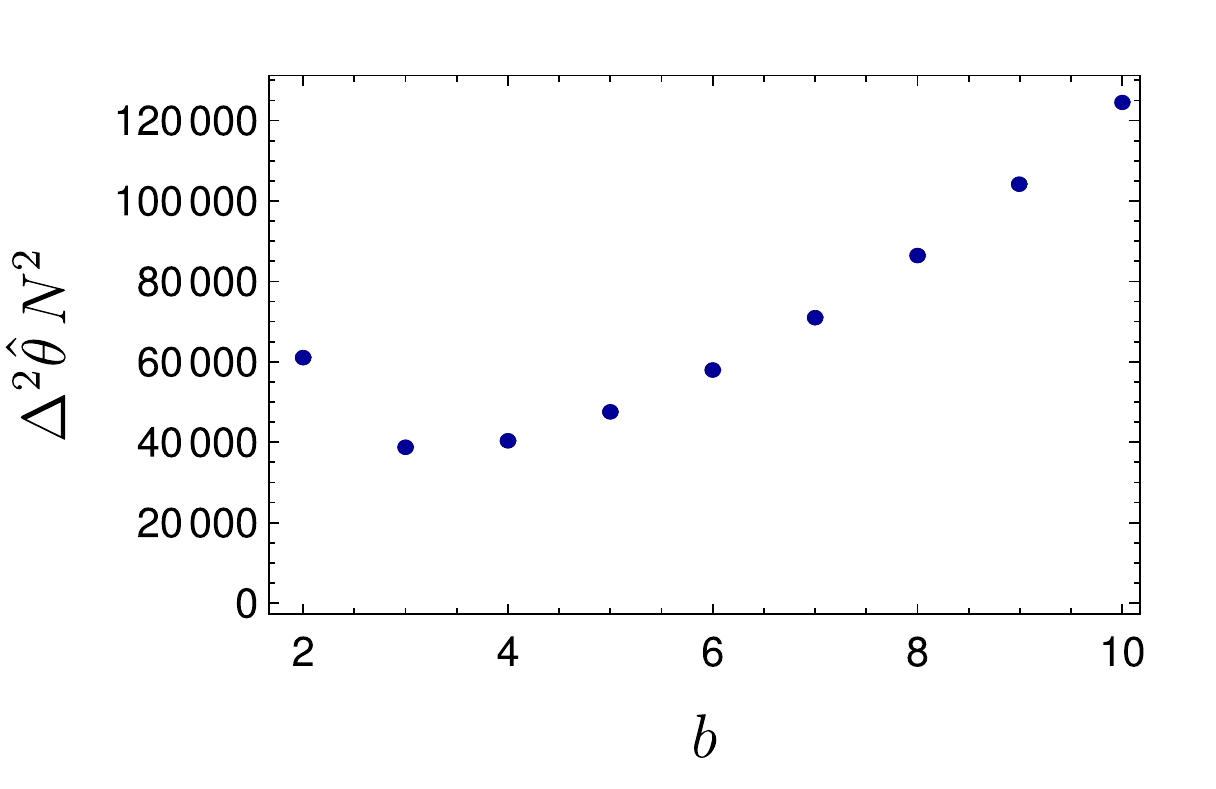}
	\end{center}
	\caption{Upper bounds on the prefactor $\var N^2$ from Eq.~\eqref{eq:lastgenericb} for discrete $b = 2, 3, \dots, 10$. It shows a minimum for $b^\star = 3$.}
	\label{fig:genericb}
\end{figure}

As discussed in Sec.~\ref{subsect:presentationAlgorithm}, in presenting the phase estimation algorithm we assumed the size of the $K$ groups to vary as in Eq.~(\ref{doubling}). This choice is not mandatory and one can imagine a strategy with different sizes for the entangled states, for example $M_j = b^{j-1}$ with $b > 2$, and for some now choice of the angular confidence interval $\pi/n$. For the algorithm to be valid we ask for one and only one intersection of each old interval around $\hat{\theta}$ with the new intervals, just as it holds in Fig.~\ref{fig:circle}, which in the present case means
\begin{equation}
	\frac{2 \pi}{n b} + \frac{2 \pi}{n} = \frac{2 \pi}{b} \qquad \Longrightarrow\qquad n=b+1\;,
	\label{eq:conditionIntersection}
\end{equation}
see Fig.~\ref{fig:line2} (of course analogous equations $\frac{2 \pi}{n b^{j}} + \frac{2 \pi}{n b^{j-1}} = \frac{2 \pi}{b^{j}}$ must hold for each $j$, but they all reduce to Eq.~\eqref{eq:conditionIntersection}). We observe that while our original choice ($b = 2$, $n = 3$) fulfills (\ref{eq:conditionIntersection}) this is not the case for the generalization of \cite{Higgins2009} presented in~\cite{Kimmel2015}, paving the way for an underestimation of the associated MSE.
Following the derivation presented in the main text we now proceed in computing the upper bound for the MSE associated with choices of $n$ and $b$ that satisfies (\ref{eq:conditionIntersection}). First of all we notice that the probability bound of Eq.~\eqref{eq:expBound} becomes:
\begin{equation}
	P \left( | \widehat{M_j \theta} - M_j \theta | \ge \frac{\pi}{n} \right) \le A C^{-\nu_j} \;,
\end{equation}
and by virtue of~\cite{Berg2019} we have $C = \exp \left[ \frac{1}{4} \sin^2 \left( \frac{\pi}{n}\right) \right]$ . The variable $A$ keeps its value $A = 4$. The optimal bound on the MSE can be derived from the Lagrangian
\begin{multline}
	\mathcal{L} = \left( \frac{\pi}{n \cdot b^{K-1}} \right)^2 + \sum_{j=1}^K \left( \frac{2 \pi }{n \cdot b^{j-2}} \frac{b}{b-1}\right)^2 A C^{-x_j} + \\ +\lambda \left(2 \sum_{j=1}^K x_j b^{j-1} \right),
\end{multline}
which accounts for the drift via the term $\frac{b}{b-1}$. From this we find
\begin{equation}
	\var \le \left( \frac{2 \pi}{n} \right)^2 \left[ \frac{b^2}{4} + \frac{b^7 A}{\left( b-1 \right)^3 C^{x_k - \frac{1}{2}}} \right] \frac{1}{b^{2 K}} \, ,
\end{equation}
and the resource upper bound of Eq.~\eqref{eq:totalResource} reads
\begin{equation}
	N_K^>:= \frac{2 b^K}{b-1} \left( \frac{\gamma}{b-1} + x_K + \frac{1}{2} \right) \, ,
\end{equation}
with $\gamma := \frac{3}{\log_b C} = \frac{12 \log b}{\sin^2 \left( \frac{\pi}{n} \right)}$. Putting this two together we get the bound on the prefactor, analogous to Eq.~\eqref{eq:last}, i.e.
\begin{multline}
	\var N^2 \le \frac{16 \pi^2}{\left( b^2 - 1 \right)^2} \left[ \frac{b^2}{4} + \frac{b^7 A}{\left( b-1 \right)^2 C^{x_k - \frac{1}{2}}} \right] \cdot \\ \cdot \left[ \frac{\gamma}{b-1} + x_K + \frac{1}{2} \right]^2 \, .
	\label{eq:lastgenericb}
\end{multline}
The idea will be to establish which $b \in \mathbb{N}$ with $b \ge 2$ is optimal regarding this bound. This analysis can be carried out by computing numerically the optimal $x_K$ as a function of $b$, and inserting it back into Eq.~\eqref{eq:lastgenericb}. This produces Fig.~\ref{fig:genericb}, which shows a minimum for $b^\star = 3$. The corresponding upper bound on the prefactor is
\begin{equation}
	\var N^2 \le \left( 62.7 \pi \right)^2 \, .
\end{equation}
Remember that this analysis is based only on an upper bound on the precision and an analytical estimation of the error probability. Neither of these are expected to be tight, nevertheless this result may suggest that the real optimal $b$ is greater than $2$.

\section{Proof of Theorem \ref{thm:optimalDeltaM}}
\label{app:proof}
To prove the statement we define a set of four moves to be applied in order to transform a distribution $\Delta \nu_j$ into another distribution $\Delta \nu_j'$ with a MSE strictly lower than that of $\Delta \nu_j$. In the end the only distribution that cannot be further lowered will be $\Delta \nu_j = b_j$, which also satisfies $\Delta \nu_j > - \round{ \gamma \left( K - j \right) + x_K }$ being $b_j \ge 0$. The first two rules are:
\begin{enumerate}
	\item If $\Delta \nu_j \ge \Delta \nu_{j+1} + 2$ then fuse a pair probes of size $2^{j-1}$ to produce a probe of size $2^{j}$. \\
	\item If $\Delta \nu_{j+1} \ge \Delta \nu_{j} + 2$ then split a probe of size $2^{j}$ into two probes of size $2^{j-1}$.
\end{enumerate}
Assuming that the above moves have been applied wherever is possible, then the difference between to consecutive $\Delta \nu_j$ can be only $\pm 1$ or $0$. The following two moves are applied under this hypothesis.
\begin{enumerate}
	\setcounter{enumi}{2}
	\item If $\exists \, l \, | \, \Delta \nu_l > 1$ then there must exist $\nu_j = 2$ for some $j$. A string containing the first occurrence (from the right) of $\Delta \nu_j = 2$ reads $2 \, 1 \, 1 \, \dots 1 \, 0$ with a certain number of ones in between $2$ and $0$. We are guaranteed to find a $0$ on the right because if $\Delta \nu_K \ge 1$ then we would have more resources than allowed ($\Delta N \le 2^{K+1} - 2$). The move is then
	\begin{equation}
	2 \, 1 \, \cdots \, 1 0 \rightarrow 0 \, 0 \, \cdots 0 \, 1 \;,
	\label{eq:move3}
	\end{equation}
	\item If some $\Delta \nu_j < 0$ then they can't be all $ < 0$, because $2 \sum_{j=1}^{K} \Delta \nu_j 2^{j-1} = \Delta N \ge 0$. Starting from the right the first $-1$ encountered must belong to a sequence of the form $-1 \, 0 \cdots 0 \, 1$ or $1 \, 0 \, \cdots 0 \, -1$ ($\Delta \nu_j$ must cross the zero). The possibility that the first $-1$ belongs to a sequence of the second kind without belonging also to a sequence of the first kind is again excluded by the requirement $\Delta N \ge 0$. The move is then
	\begin{equation}
	-1 \, 0 \cdots 0 \, 1 \rightarrow 1 \, 1 \, \cdots 1 \, 0 \;.
	\label{eq:move4}
	\end{equation}
\end{enumerate}
The idea is that after the application of one of the last two moves we have to apply wherever possible the first two before applying again 3 or 4. A distribution allowing the above moves can't be a minimizer because we can modify it to have a strictly lower MSE. Therefore the minimizer must be searched among the distributions to which the moves don't apply, which are those with $\Delta \nu_j \in \lbrace 0, 1 \rbrace$. There happens to be only one of such distributions which is the binary writing of $\Delta N$. We now show that each of the four moves gives a decrease in the MSE bound.
\begin{enumerate}
	\item Let's confront the changes in the summation on the right hand of bound~\eqref{eq:modifiedSummation} before and after the first move, i.e.
	\begin{align}
		\begin{split}
			 2^{j - \log_2 C \Delta \nu_j} + 2^{j+1 - \log_2 C \Delta \nu_{j+1}} > \\ 2^{j-\log_2 C \left( \Delta \nu_j - 2 \right)} + 2^{j+1 - \log_2 C \left( \Delta \nu_{j+1} + 1\right)} \;,
		\end{split}	
	\end{align}
	this means
	\begin{align}
		\begin{split}
			\quad \quad &\Delta \nu_j > \Delta \nu_{j+1} + \log_2 \left( \frac{C^2 + C}{2} \right) \Big / \log_2 C \Longrightarrow \\ \quad \quad &\Delta \nu_j \ge \Delta \nu_{j+1} + 2 \;.
		\end{split}
	\label{eq:comparisonMove1}
	\end{align}
	The last implication holds because $\nu_K$ and $\nu_{K+1}$ are integers and $C > 1$. In this case there is always a non zero gap between the MSE before and after the application of the rule.
	\item We now determine when the reverse move of splitting a probe is useful.
	\begin{align}
		\begin{split}
			\quad 2^{j-\log_2 C \left( \Delta \nu_j + 2 \right)}+2^{j+1 - \log_2 C \left( \Delta \nu_{j+1}-1 \right)} \\ < 2^{j- \log_2 C \Delta \nu_j} + 2^{j+1 - \log_2 C \Delta \nu_{j+1}} \;,
		\end{split}
	\end{align}
	that is
	\begin{align}
		\begin{split}
			\quad \quad &\Delta \nu_{j+1} > \Delta \nu_j + \log_2 \left( \frac{2 C^2}{1 + C} \right) \Big / \log_2 C \;, \\
		\end{split}
	\label{eq:comparisonMove2}
	\end{align}
	so the splitting is convenient if $\Delta \nu_{j+1} \ge \Delta \nu_{j} + 2$ and also in this case the gap between the MSE before and after the application of the rule is always positive.
	\item Let's say that in rule~\eqref{eq:move3} there are $\alpha - 1$ ones in the middle of the left hand side. We compare the MSE before and after the move only for the affected part of the summation in bound~\eqref{eq:modifiedSummation} (regardless of common factors).
	\begin{align}
		\begin{split}
			\quad 2^{- 2 \log_2 C} + 2^{1 - \log_2 C } + \dots + 2^{\alpha - 1 - \log_2 C} + 2^{\alpha} \\ > 1 + 2 + \dots + 2^{\alpha - 1} + 2^{\alpha - \log_2 C} \;,
		\end{split}
	\end{align}
	that means
	\begin{align}
		\begin{split}
			\quad C^{-2} + C^{-1} \left( 2^{\alpha} - 2 \right) + 2^{\alpha} > 2^{\alpha} - 1 + 2^{\alpha} C^{-1} \;,
		\end{split}
	\end{align}
	this last inequality implies $\left( C - 1 \right)^{2} > 0$. Therefore also in this case we have a finite gap and it is convenient to perform the move.
	\item Let's say that in rule~\eqref{eq:move4} there are $\alpha - 1$ ones in the middle of the right hand side. Then the comparison between the MSE bounds reads (neglecting common factors)
	\begin{align}
		\begin{split}
			\quad 2^{ - \log_2 C} + 2^{1 - \log_2 C} + \cdots 2^{\alpha - 1 - \log_2 C} + 2^{\alpha} \\ < 2^{\log_2 C} + 2^{1} + \dots + 2^{\alpha - 1} + 2^{\alpha- \log_2 C} \;,
		\end{split}
	\end{align}
	which is
	\begin{align}
		\begin{split}
			\quad C^{-1} \left( 2^{\alpha} - 1 \right) + 2^{\alpha} < C + \left( 2^{\alpha} - 2 \right) + C^{-1} 2^{\alpha} \;,
		\end{split}
	\end{align}
	again the last inequality is $\left( C - 1 \right)^2 > 0$ and there is a positive decrease of the MSE when the move is performed.
\end{enumerate}
This closes the proof of Theorem \ref{thm:optimalDeltaM}.

\section{Upper bound for odd $\Delta N$}
\label{app:smallSteps}
Theorem~\ref{thm:optimalDeltaM} states that the number of extra probes is of the form $\Delta N = 2 \sum_{j = 1}^{K} b_j 2^{j-1}$, therefore it must be even. This stems from the fact that the algorithm assumes that an equal number of measurements $\nu_j$ is performed for every step of the estimation. Even the single probe measurement at step $j = 1$ requires the resources to be evenly distributed between measurements of Type-$0$ and of Type-$+$. Suppose that we are given an extra probe, then it may be used to enhance one of the measurements at step $j = 1$. Let's say without loss of generality that measurement of Type-$0$ is now performed with $\nu_1 + 1$ probes (Type-$+$ still employs $\nu_1$ measurements), then the probability bound~\eqref{eq:prob1} becomes:
\begin{eqnarray}
	\text{P} \left( | f_0 - p_0 | \ge \varepsilon \right) &\le& 2 \exp \left( - 2 \nu_1 \varepsilon^2 - 2 \varepsilon^2 \right) \\ &=& \frac{2}{C} \exp \left( - 2 \nu_1 \varepsilon^2 \right) \;,
\end{eqnarray}
where $C = \exp \left( 2 \varepsilon^2 \right)$. The analytical bound~\eqref{eq:anayticalBound} is modified as
\begin{equation}
	\text{P} \left( | \widehat{M_1 \theta} - M_1 \theta | \ge \frac{\pi}{3} \right) \le 4 C^{-\nu_1} \left[ 1 - \frac{1}{2} \left( 1 - \frac{1}{C} \right) \right] \; ,
\end{equation}
We assume that such modification applies to every bound of the form in Eq.~\eqref{eq:expBound}, even if it has not been derived from the Hoeffding's bound. So in general
\begin{equation}
	\text{P} \left( | \widehat{M_1 \theta} - M_1 \theta | \ge \frac{\pi}{3} \right) \le A C^{-\nu_1} \left[ 1 - \frac{1}{2} \left( 1 - \frac{1}{C} \right) \right] \; .
\end{equation}
We start again from Eq.~\eqref{eq:RMSEBound} and add a probe to the first step, by modifying the error probability as prescribed the bound becomes
\begin{multline}
	\var \le \left( \frac{2 \pi}{3} \right)^2 \left( \frac{1}{4^K} + 16 \sum_{j = 1}^{K} \frac{A}{4^{j-1}} C^{-\nu_j} \right) \\ - \left( \frac{2 \pi}{3} \right)^2 \left( 1 - \frac{1}{C} \right) \frac{64 A} {2^{3 K} C^{x_K - \frac{1}{2}}} \;.
	\label{eq:oddBound}
\end{multline}
The first part of this expression can be optimized as in Sec.~\ref{subsect:interpolation} to get bound~\eqref{eq:optimalRMSEK} with an even number of resources. It is legit to single out a probe from the optimization as it will have no role in the measurement scheme. The extra term in Eq.~\eqref{eq:oddBound} is the same term arising from Eq.~\eqref{eq:optimalRMSEK} by incrementing $\Delta N \rightarrow \Delta N + 1$. Therefore for an odd $\Delta N$ this procedure gives exactly bound~\eqref{eq:optimalRMSEK}, so the applicability of this formula depends no more on the parity of $\Delta N$.

\section{Optimal redistribution for negative $\Delta N$}
\label{app:optimalK}
In this appendix we answer the following question: what happens if we reduce the number of probes but we are bound to keep the same (fixed) size for the biggest entangled state? In particular, what is the optimal distribution of probes? Equivalently what is the optimal distribution when $\Delta N < 0$? Such question was not relevant to compute the distribution of resource as it is not convenient to force the input state to be more entangled than the ramp in Eq.~\eqref{eq:linearRamp} suggests. Nevertheless to answer this question we modify Eq.~\eqref{eq:total} with $\Delta \nu_j$ and write
\begin{multline}
	\var \le \left( \frac{2 \pi}{3} \right)^2 \left( \frac{1}{4^K} + \frac{64 A }{2^{3 K} C^{x_K - \frac{1}{2}}} \sum_{j=1}^{K} 2^{j-\log_2 C \cdot \Delta \nu_j} \right) \;, \nonumber 
\end{multline}
this time we have $x_{K}$ fixed ($\Delta \nu_{K} = 0$) and $\Delta \nu_j \le 0$. Each time $\Delta N = - 2 \left( 2^K - 1 \right)$ we have $\Delta \nu_j = -1$ for $j = 1, 2, \dots, K-1$, at this point we reset all the counters $\Delta \nu_j$ and $\Delta N$, accounting this contributions as a common factor in the next step , indeed $\sum_{j=1}^K 2^{j - \log_2 C \Delta \nu_j} = C\sum_{j=1}^{K} 2^j$. By changing signs to $\Delta \nu_j$ in the four moves, Theorem \ref{thm:optimalDeltaM} is still valid (with $b_j \in \lbrace 0, -1 \rbrace$), but we don't report here the necessary checks. In the end we get
\begin{eqnarray}
	\begin{aligned}
	\var &\le \left( \frac{2 \pi}{3} \right)^2 \left( 1 + \frac{128 A }{C^{x_K - \frac{1}{2}}} \right) \frac{1}{4^K} \\ \quad &- \left( \frac{2 \pi}{3} \right)^2 \frac{64 A \left( C - 1 \right) C^i}{2^{3 K} C^{x_K - \frac{1}{2}}} \Delta N \;,
	\end{aligned}
	\label{eq:optimalRMSEK1}
\end{eqnarray}
where the index $i$ start as $i = 0$ and is raised by one at every saturation of the $\Delta \nu_j$ variables. The formula in bound~\eqref{eq:optimalRMSEK1} has been obtained by noticing that
\begin{equation}
	\sum_{j=1}^{K} 2^{j - \log_2 C \cdot \Delta \nu_j} = \sum_{j=1}^{K} 2^{j} - \left( C - 1 \right) \Delta N \;.
\end{equation}
We don't use this bound in the main text as it will never be optimal in comparison to strategies with less entanglement.

\section{Adaptive measurement}
\label{app:adaptive}
In this appendix we present a manipulation which consists in applying to each probe of the codified state $|{\rm GHZ}_{\theta} ^{(M_j)}\rangle$ the phase shift $V_{\phi} := e^{-i \phi H}$, generating
\begin{equation}
	|{\rm GHZ}^{(M_j)}_{\theta-\phi} \rangle = (|0\rangle^{\otimes M_j} + e^{i M_j\left( \theta - \phi \right)} |1\rangle^{\otimes M_j}) /\sqrt{2} \;,
	\label{OUTPUTPHIreal} 
\end{equation}
In Sec.~\ref{subsect:limitedEnt}, after the $(K-1)$-th step has been successfully executed, we know the phase to be in an interval of size $\frac{1}{2^{K-2}} \frac{2 \pi}{3}$. By applying an appropriate shift operator $V_{\phi_1}$, we can make the computed interval for $\hat{\theta}$ at the $(K-1)$-th step completely contained in one of the periods of $|{\rm GHZ}^{(M_K)}_{\theta-\phi} \rangle$, with $M_K = 2^{K-1}$, being them of size $\frac{2 \pi }{2^{K-1}} > \frac{1}{2^{K-2}} \frac{2 \pi}{3}$. This resolves the period ambiguity in the last step. For each state of size $M_K$, numbered with the index $i = 1, \dots, 2 \nu_K$, a control $V_{\phi_i}^{\otimes M_K}$ is applied. Each entangled state is projected onto $(|0\rangle^{\otimes M_j} \pm |1\rangle^{\otimes M_j}) /\sqrt{2}$. This produces as outcome a Bernoulli variable with value $0$ or $1$ characterized by outcome probabilities
\begin{equation}
	p^i_0 := \frac{1 + \cos M_K (\theta-\phi_i)}{2} \;, \qquad p^i_1= 1-p^i_0\;,
\end{equation}
where each $\phi_i$ is chosen according to the previous records. The Maximum Likelihood Estimator extracted from the collected data saturates the QCR bound, Eq.~\eqref{eq:lastEstimator}, in the limit $\nu_K \rightarrow \infty$, see~\cite{Fujiwara2006}, when suitable $\phi_i$ are chosen. Again the above detection procedure can be implemented via local detection of the individual probes of~(\ref{OUTPUTPHI}) -- see Ref.~\cite{Bollinger1996} and the discussion presented in Appendix~\ref{app:SEPMES}. Notice that by keeping the error interval $\frac{8 \pi}{3 \cdot 2^{j-1}}$ in Eq.~\eqref{eq:entLimLagrangian}, as explained in Sec.~\ref{subsect:analysisAlgorithm}, we account for the possible accumulation of errors also in this modified strategy.

\end{document}